# The Disintegration of Free Speech

Yiyang Mei[*]

## Abstract

Free speech in the United States is facing a crisis. Since President Trump's return to office, the federal government has invoked the rhetoric of restoration while deploying institutional power to discipline disfavored expression. At the same time, the communicative environment has undergone a structural transformation. Social media platforms, increasingly integrated with generative AI systems, now function as the primary sites of public communication. Within this environment, AI-generated pornography and large-scale political misinformation have produced widely documented dignitary and democratic harms. In response, states across the political spectrum have enacted laws aimed at preserving democratic order, curbing AI-mediated harms, and promoting safer online environments. Yet these regulatory efforts run headlong into the Free Speech Clause of the First Amendment. Requirements that platforms remove content, disclose recommendation practices, or redesign moderation systems are readily characterized as compelled speech or as interference with protected editorial judgment. This Article argues that, under prevailing First Amendment doctrine, AI-generated content remains protected speech notwithstanding its harms, and that regulations targeting platform content moderation are likely unconstitutional. Since the 1970s, the Supreme Court has shifted from protecting the free circulation of information to privileging editorial autonomy, understood as control over authorship, expressive identity, and freedom from state attribution. Once content moderation is treated as an exercise of editorial judgment, regulatory mandates that compel or prohibit such practices presumptively violate the Free Speech Clause. The Article concludes that this doctrinal trajectory risks detaching the First Amendment from its democratic purposes and calls for a reconstruction capable of addressing contemporary communicative infrastructures, automated production, and concentrated communicative power.

---

[**] Yiyang Mei, J.D., M.P.H., S.J.D. Candidate (2026). The author is grateful to Professor John Witte, Jr. for his guidance and supervision of this project.

**Table of Contents**

**INTRODUCTION**

Anyone seriously concerned with free expression should be alarmed by the present condition of this foundational liberty in the United States. Since President Trump's return to office in January 2025, the administration has framed its speech agenda in the rhetoric of restoration. On his first day back in office, the President issued an executive order titled *Restoring Freedom of Speech and Ending Federal Censorship*, accusing the prior administration of suppressing disfavored viewpoints and pledging to secure constitutionally protected expression and remedy past governmental misconduct.[1] The actions that followed, however, cut sharply in the opposite direction. The administration barred the Associated Press from press events for refusing to adopt preferred geographic terminology,[2] initiated deportation proceedings against lawful immigrants for online speech,[3] pursued libel actions against major newspapers such as *The Wall Street Journal*,[4] moved to dismantle Voice of America for alleged ideological nonconformity,[5] and conditioned National Endowment for the Arts funding on the rejection of projects deemed to "promote gender ideology."[6] Taken together, these measures reflect a state willing to deploy institutional and legal power to discipline speech it finds objectionable, even while professing fidelity to the First Amendment.

At the same time, the communicative environment itself is undergoing a structural transformation. Social media platforms now function, as Justice Alito observed in *Moody v. NetChoice, LLC*, as the "modern public squares," spaces without which life, particularly for younger users, is scarcely imaginable.[7] These platforms have become the primary means through which individuals communicate with family and friends, conduct everyday transactions, engage in commerce, and learn about or comment on current events.[8] Increasingly embedded within these platforms are large language model based chatbots and other generative AI services. Since their widespread introduction into daily life in late 2022, these systems have been used to form emotionally intimate relationships,[9] generate nonconsensual and sexually explicit images,[10] and fabricate and disseminate political misinformation at unprecedented scale and speed, including viral falsehoods involving public figures and elections.[11] The resulting harms are well documented. A growing body of reporting in popular media and major news outlets have drawn sustained attention to dignitary harms arising from distorted self conception,[12] the exploitation of vulnerability,[13] asymmetric power relationships,[14] and intrusions on privacy,[15] as well as to systemic harms to electoral integrity,[16] institutional trust,[17] and democratic order.[18] These effects present a self-evident case for regulatory intervention.

---

[1] *See Restoring freedom of Speech And Ending Federal Censorship,* THE WHITE HOUSE (January 20, 2025).
[2] *See* David Bauder, *White House Says It Has the Right to Punish AP Reporters over Gulf Naming Dispute*, AP, https://apnews.com/article/ap-white-house-gulf-name-dispute-3f43c519a4b4f4661dd0831421943ef7 (last updated February 12, 2025).
[3] *See* Erin Corcoran, *Lawful Permanent Residents like Mahmoud Khalil Have a Right to Freedom of Speech - But Does That Protect Them from Deportation?*, THE CONVERSATION (April 21, 2025), https://theconversation.com/lawful-permanent-residents-like-mahmoud-khalil-have-a-right-to-freedom-of-speech-but-does-that-protect-them-from-deportation-254042
[4] *See* Peter Charalambous, *Trump Files $15 Billion Defamation Suit Against New York Times, Penguin Random House*, ABC NEWS (September 16, 2025), https://abcnews.go.com/US/trump-files-15-billion-defamation-suit-new-york/story?id=125619941.
[5] *See* Thomas Mackintosh, Merlyn Thomas, *Trump Moves to Close Down Voice of America*, BBC (March 16, 2025), https://www.bbc.com/news/articles/cvge4l109r3o
[6] *See* Elizabeth Blair, *"Chilling Effect": Arts Organizations React to End of DEI Initiatives from Fed Agency*, NPR (February 11, 2025), https://www.npr.org/2025/02/11/nx-s1-5293082/trump-executive-orders-dei-nea-arts-organizations
[7] 603 U.S. 707, 767 (2024).
[8] *Id.*
[9] *See* Efua Andoh, *Many Teens are Turning to AI Chatbots for Friendship and Emotional Support*, AMERICAN PSYCHOLOGICAL ASSOCIATION (October 1, 2025), https://www.apa.org/monitor/2025/10/technology-youth-friendships
[10] *See Generative AI & Sexually Explicit Deepfakes*, NORTH DAKOTA DOMESTIC & SEXUAL VIOLENCE COALITION (April 9, 2025), https://nddsvc.org/generative-ai-sexually-explicit-deepfakes.

In response to these developments, states have enacted a range of laws aimed at preserving democratic order, curbing AI mediated harms, and promoting safer online environments. These regulatory efforts track familiar ideological divides.[19] States such as Florida and Texas have adopted so-called anti-censorship regimes that prohibit platforms from deplatforming, deprioritizing, or otherwise discriminating against content on the basis of viewpoint.[20] By contrast, states such as California and New York have pursued affirmative regulatory strategies, imposing duties on platforms to engage in content moderation, mandating disclosures regarding algorithmic ranking, recommendation, and training practices, requiring heightened protections where children are likely to be exposed to manipulative or harmful interface designs, banning deceptive chatbot interactions, and establishing takedown obligations for nonconsensual intimate imagery.[21] Across the political spectrum, states have also enacted measures targeting practices that distort or interfere with electoral processes.[22]

---

[11] *See* Nicol Turner Lee, Darrell M. West, and Kathryn Dunn Tenpas, *How Do Artificial Intelligence and Disinformation Impact Elections*, BROOKINGS (October 10, 2024), https://www.brookings.edu/articles/how-do-artificial-intelligence-and-disinformation-impact-elections/
[12] *See* Marlynn Wei, *The Emerging Problem of "AI Psychosis"*, PSYCHOLOGY TODAY (November 27, 2025), https://www.psychologytoday.com/us/blog/urban-survival/202507/the-emerging-problem-of-ai-psychosis (explaining that correspondence with generative AI chatbots such as ChatGPT results in cognitive dissonance, which may fuel delusions in those with increased propensity toward psychosis).
[13] *See* Abdul-Fatawu Abdulai, *Is Generative AI Increasing the Risk for Technology-Mediated Trauma Among Vulnerable Populations?*, 32(1) NURSING INQUIRY e12686 (2024).
[14] *See* John Nosta, *The Precarious Asymmetries of Human-AI Relationships*, PSYCHOLOGY TODAY (April 1, 2024), https://www.psychologytoday.com/us/blog/the-digital-self/202404/the-precarious-asymmetries-of-human-ai-relationships,
[15] *See Privacy in an AI Era: How Do We Protect Our Personal Information*, HAI (March 18, 2024), https://hai.stanford.edu/news/privacy-ai-era-how-do-we-protect-our-personal-information
[16] *See* Tal Feldman and Aneesh Pappu, *The Era of AI Persuasion in Elections is About to Begin*, MIT TECHNOLOGY REVIEW (December 5, 2025), https://www.technologyreview.com/2025/12/05/1128837/the-era-of-ai-persuasion-in-elections-is-about-to-begin/; *See also* Deni Ellis Bechard, *AI Chatbots are Shockingly Good at Political Persuasion*, SCIENTIFIC AMERICAN (December 4, 2025), https://www.scientificamerican.com/article/ai-chatbots-shown-to-sway-voters-raising-new-fears-about-election-influence/.
[17] *See* Ivan Shkvarun, *Trust is the New Currency in the AI Agent Economy*, WORLD ECONOMIC FORUM (Jul 25, 2025), https://www.weforum.org/stories/2025/07/ai-agent-economy-trust/; Alexis Bonnell, *Navigating the Paradox: Restoring Trust in an Era of AI and Distrust*, NATIONAL ACADEMY OF PUBLIC ADMINISTRATION (September 22, 2023), https://napawash.org/standing-panel-blog/navigating-the-paradox-restoring-trust-in-an-era-of-ai-and-distrust
[18] *See Raluca Csernatoni*, *Can Democracy Survive the Disruptive Power of AI*, CARNEGIE ENDOWMENT FOR INTERNATIONAL PEACE (December 18, 2024), https://carnegieendowment.org/research/2024/12/can-democracy-survive-the-disruptive-power-of-ai?lang=en); *see also* Alla Polishchuk, *AI Poses Risks to Both Authoritarian and Democratic Politics*, WILSON CENTER (January 26, 2024).
[19] *See infra* Part 1, section B.
[20] *Id.*
[21] *Id.*
[22] *Id.*

These regulatory efforts, however, run headlong into the Free Speech Clause of the First Amendment. Requirements that platforms remove content, explain moderation decisions, disclose recommendation practices, or redesign their systems to mitigate public harms plausibly appear as compelled speech or as direct interference with protected editorial judgment. New York's Hateful Conduct Law, for example, by requiring platforms to receive and respond to user complaints concerning "hateful" expression, can be analogized to compelled affirmation of state defined moral judgments, much like the forced display of the state motto invalidated in *Wooley v. Maynard*.[23] Its SAFE for Kids Act raises parallel concerns. By mandating disclosures regarding recommendation system logic and prescribing how platforms must present and prioritize information,[24] the statute risks governmental control over platforms' own expressive choices, a form of interference the Court rejected in *Pacific Gas & Electric Co. v. Public Utilities Commission of California*.[25] California's Age Appropriate Design Code Act likewise presents similar First Amendment concerns and is currently under review in *NetChoice v. Bonta* before the Ninth Circuit.[26] That law limits how platforms may collect, use, and design around children's data and requires data protection impact assessments, prompting challenges that it restricts the availability and use of information and compels speech in ways condemned in *Sorrell v. IMS Health*.[27] There, the Court struck down Vermont's Prescription Confidentiality Law for imposing content and speaker based burdens on the dissemination of truthful information,[28] drawing on a line of cases protecting factual expression ranging from credit reports[29] to product labeling[30] and lawfully obtained communications.

Within this context, the First Amendment, originally promulgated to secure individual self fulfillment,[31] the pursuit of truth,[32] and meaningful participation in public life,[33] is increasingly invoked to erect constitutional obstacles to regulations aimed at serving those same ends. The questions that follow are therefore unavoidable: should such challenges prevail; does the First Amendment permit states to regulate platforms' decisions to curate, rank, suppress, or explain content; and does that protection extend to AI generated expression?

This Article argues that, under prevailing First Amendment doctrine, AI generated content, notwithstanding its documented harms, is protected speech. It further argues that state regulations targeting platform content moderation are likely to be held unconstitutional under the First Amendment. Since the 1970s, the Court has shifted from protecting the free circulation of information to privileging editorial autonomy, understood as control over authorship, subject matter, expressive identity, and freedom from compelled or state attributed speech. Once moderation practices such as curating, ranking, suppressing, labeling, contextualizing, or refusing content are understood as exercises of editorial judgment, governmental efforts to compel, prohibit, or restructure those choices presumptively violate the Free Speech Clause.

---

[23] *See* 430 U.S. 705 (1977).
[24] S. 7694-A, 2023-2024 Reg. Sess. (N.Y. 2023).
[25] 475 U.S. 1 (1986).
[26] 152 F. 4th 1002 (9th Cir. 2025).
[27] 564 U.S. 552 (2011).
[28] *Id.*
[29] *See e.g.,* Dun & Bradstreet, Inc. v. Greenmoss builders, Inc., 472 U.S. 749, 759 (1985) (holding credit report is "speech").
[30] *See e.g.,* Rubin v. Coors Brewing Co., 514 U.S. 476, 481 (1995) ("information on beer labels" is speech).
[31] *See* Thomas I. Emerson, *Toward a General Theory of the First Amendment*, 72 YALE L.J. 877, 879-881 (1963).
[32] *Id* at 881-882.
[33] *Id* at 882-884.

To advance this argument, this Article proceeds in three Parts. Part I sets the context for the constitutional analysis that follows by identifying two categories of AI-mediated harm that have prompted urgent regulatory attention: nonconsensual sexually explicit generative content and AI-generated political misinformation. It then surveys the content-moderation laws that states across the political spectrum have enacted in response, framing those interventions as efforts to preserve individual dignity and secure democratic order. Part II argues that, in spite of the gravity of these harms, both categories of content remain largely protected by the First Amendment under existing doctrine. The analysis includes four steps: it first addresses the threshold question whether the Free Speech Clause applies to AI-generated content at all, then evaluates the applicability of the established categorical exceptions for obscenity, hate speech, and fighting words, concluding that none comfortably addresses the content at issue. Part III turns to the regulation of platforms themselves, arguing that content-moderation mandates, however well-intentioned, are likely to be found unconstitutional insofar as existing doctrine treats moderation decisions as exercises of protected editorial judgment. Methodologically, this Part uses a combination of close reading of the Free Speech Clause, informed by Founding-era dictionaries, followed by a systematic reconstruction of First Amendment case law from the 1940s to the present.

The implication for this Article is that the Free Speech Clause may be approaching a moment of doctrinal disintegration. When First Amendment doctrine is repeatedly, consistently, and increasingly successfully, deployed to invalidate regulations aimed at preserving individual autonomy, protecting children from manipulative and exploitative systems, safeguarding democratic processes from large-scale deception, and maintaining a minimally safe communicative environment, it becomes untethered from the institutional conditions that once gave its protections public meaning. In doing so, it ceases to function as a framework for democratic self-government and instead operates as a generalized veto on governance itself. If the Clause is to continue serving its original purpose of securing public participation in a liberal democracy, its core commitments must be reconstructed into a coherent theory capable of addressing speech infrastructures, automated production, and concentrated communicative power, rather than remaining confined to a binary opposition between state censorship and private autonomy.

## PART I. SOCIAL HARMS AND THEIR CORRESPONDING REGULATIONS

This Part sets the context for the constitutional analysis that follows by identifying two categories of AI-mediated harm that have prompted regulatory intervention: harms arising from AI-generated content and harms arising from the dissemination of misinformation on digital platforms. It then surveys the regulatory responses that states across the political spectrum have adopted to address those harms.

## A. Two Harms by AI

This section identifies two paradigmatic categories of AI-mediated harm: nonconsensual sexually explicit generative content, and AI-generated political misinformation that destabilizes markets, undermines institutional trust, and corrodes democratic order. Both present compelling cases for regulatory intervention aimed at preserving individual dignity and the conditions of democratic self-government.

### 1. AI-Generated Pornography and Chatbots Companions

The first category of harm involves AI-generated explicit images and fictional characters designed to function as emotional companions, fostering forms of dependency and intimacy among children and teenagers that, in documented cases, have culminated in self-harm and suicide. For the AI-generated explicit images, as of March 2024, barely two years after the introduction of generative AI systems, nearly 4,000 celebrities, including female actors, television personalities, musicians, and YouTubers, had been identified as victims of deepfake pornography, with their faces superimposed onto explicit material.[34] Ordinary individuals fared no better. An eighteen-year-old Australian law student, Noelle Martin, for example, discovered explicit images of herself posted on pornographic websites, in which her face had been doctored onto the naked bodies of adult actresses.[35] This represents a dramatic escalation of an already pervasive phenomenon - since 2018, the volume of pornographic deepfakes has doubled approximately every six months.[36] By 2021, more than 180,000 deepfake videos had been posted online.[37] That figure rose to approximately 720,000 by the summer of 2022, generating hundreds of millions of views.[38]

This scale of production has been further accelerated by the emergence of platforms explicitly designed to lower technical barriers. Rather than requiring thousands of images and hours of video footage as before, many applications now allow users to generate deepfakes from a single photograph, often scraped directly from a victim's social media account.[39] Once an image is uploaded, users can select from a library of pornographic videos and generate a face-swapped preview within seconds.[40] Other applications go further, enabling users to digitally remove clothing from images of women and produce realistic synthetic nudes.[41]

---

[34] *See* Nadeem Badshan, *Nearly 4,000 Celebrities Found to be Victims of Deepfake Pornography*, THE GUARDIAN (March 21, 2024), https://www.theguardian.com/technology/2024/mar/21/celebrities-victims-of-deepfake-pornography.

[35] *See* Justin Sherman, *"Completely Horrifying, Dehumanizing, Degrading": One Woman's Fight Against Deepfake Porn*, CBS NEWS, (October 14, 2021), https://www.cbsnews.com/news/deepfake-porn-woman-fights-online-abuse-cbsn-originals/

[36] *See* Deeptrace, THE STATE OF DEEPFAKES: LANDSCAPE, THREATS, AND IMPACT 1 (2019); Nina Schick, *Deepfakes Are Jumping from Porn to Politics. It's Time to Fight Back*, WIRED (Dec. 28, 2020), https://www.wired.com/story/deepfakes-porn-politics/.

[37] *Id.*

[38] *Id.*

[39] *See* Karen Hao, *A Horrifying New AI App Swaps Women into Porn Videos with a Click*, MIT TECH. REV. (Sept. 13, 2021), https://www.technologyreview.com/2021/09/13/1035449/ai-deepfake-app-face- swaps-women-into-porn/.

[40] *Id.*

[41] One such app is Lensa. It allows users to generate hyper-realistic avatars of themselves after uploading at least 10 selfies; the generated content includes nude and hypersexualized images of the users. *See* Zoe Sottile, *What to Know about Lensa, the AI Portrait App All Over Social Media*, CNN (July 26, 2023), https://www.cnn.com/style/article/lensa-ai-app-art-explainer-

Beyond AI-generated pornographic images, AI-generated fictional characters have likewise induced children and teenagers to form intimate relationships with chatbots. The BBC has reported that Meta's chatbots were programmed to engage in explicit and "sensual" conversations with users as young as eight years old.[42] In 2023, Snapchat introduced "My AI," a virtual companion designed to learn user preferences through ongoing interaction.[43] By September of that year, Google Trends data showed a 2,400 percent increase in searches for "AI girlfriends."[44] Millions of users now turn to chatbots for emotional support and erotic roleplay.[45] Multiple reports and complaints document cases in which teenagers attempted or committed suicide after developing intense emotional attachments to fictional AI characters.[46] A complaint filed on behalf of Juliana Peralta, a thirteen-year-old in Colorado, described extensive chatbot interactions involving "hypersexual conversations that, in any other circumstance and given Juliana's age, would have triggered criminal investigation."[47] In a separate case, the family of a girl identified as "Nina" from New York alleged that their daughter attempted suicide after her parents sought to cut off her access to Character.AI.[48] In the weeks preceding the attempt, the chatbots allegedly engaged in sexually explicit roleplay, manipulated her emotional state, and fostered a false sense of intimacy and dependence.[49]

The effects of misuse of generative AI technologies involving nonconsensual and child pornography are sufficiently self-evident that they require little elaboration. A substantial body of literature has documented psychological harms,[50] including the infliction of "extreme emotional distress,"[51] the production of "feelings of low self-esteem or worthlessness," and the social isolation of victims.[52] These practices also impose broader societal harms such as eroding trust in institutions, exacerbating

---

trnd; Melissa Heikkilä, *The Viral AI Avatar App Lensa Undressed Me—Without My Consent*, MIT TECH. REV. (Dec. 12, 2022), https://www.technologyreview.com/2022/12/12/1064751/the-viral-ai-avatar-app-lensa-undressed-me-without-my-consent/.

[42] *See* Charlotte Edwards, *Meta Investigated over AI Having "Sensual" Chats with Children*, BBC (August 18, 2025), https://www.bbc.com/news/articles/c3dpmlvx1k2o.

[43] *See* James Muldoon, *"Maybe We Can Role-Play Something Fun": When an AI Companion Wants Something More*, BBC https://www.bbc.com/future/article/20241008-the-troubling-future-of-ai-relationships.

[44] *Id.*

[45] *Id.*

[46] *See Social Media Victims Law Center Files Three New Lawsuits on Behalf of Children Who Died of Suicide or Suffered Sex Abuse by Character.AI*, SOCIAL MEDIA VICTIMS LAW CENTER (September 16, 2025), https://socialmediavictims.org/press-releases/social-media-victims-law-center-files-three-new-lawsuits-on-behalf-of-children-who-died-of-suicide-or-suffered-sex-abuse-by-character-ai/.

[47] *See* Hadas Gold, *More Families Sue Character.AI developer, Alleging App Played a Role in Teens' Suicide and Suicide Attempt*, CNN (Sept 16, 2025), https://www.cnn.com/2025/09/16/tech/character-ai-developer-lawsuit-teens-suicide-and-suicide-attempt.

[48] *Id.*

[49] *Id.*

[50] *See* Allen Frances, Luciana Ramos, *Preliminary Report on Dangers of AI Chatbots*, PSYCHIATRIC TIMES, https://www.psychiatrictimes.com/view/preliminary-report-on-dangers-of-ai-chatbots (October 7, 2025).

[51] *See* State v. VanBuren, 2018 VT 95, 214 A.3d 791, 810.

[52] *See* People v. Austin, 155 N.E.3d 439, 461 (Ill. 2019). *See also* Ivan Taylor, *Megan Thee Stallion Testifies in Miami Defamation Trial Over Alleged AI Deepfake Video*, CBS NEWS (November 21, 2025), https://www.cbsnews.com/miami/news/megan-thee-stallion-testifies-miami-defamation-trial-alleged-ai-deepfake-video/.

social divisions, and undermining public safety.[53] Taken together, they present an urgent and compelling case for regulating such online and chatbot-mediated behavior.

**2. Political Misinformation**

The second category of harm involves political misinformation. A substantial body of empirical literature now establishes that digital platforms systematically amplify political mis- and disinformation. As early as 2018, Vosoughi et al. analyzed approximately 126,000 Twitter information cascades and found that false news diffused "significantly farther, faster, deeper, and more broadly" than true news across all topical domains, largely because false stories were more novel and emotionally evocative, inducing reactions such as surprise or disgust that increased users' propensity to share them.[54] Subsequent work reinforced these findings. Shao et al. (2018), examining more than fourteen million tweets linking to hundreds of thousands of low-credibility news articles during the 2016 U.S. election, showed that social bots played a disproportionate role in early-stage dissemination by aggressively seeding false content, including through replies and mentions directed at influential users, thereby jumpstarting viral momentum later sustained by unwitting human resharing.[55] Guess et al. (2019), tracking Facebook activity during the same election cycle, similarly found that a small fraction of users accounted for the majority of fake-news shares, with adults over sixty-five sharing false news at several times the rate of younger users even after controlling for ideology and partisanship.[56] Against this backdrop, and given that as of April 2025 approximately 5.64 billion individuals were active internet users,[57] 5.31 billion social media accounts were in use globally, and average daily social media use exceeded two hours,[58] even brief bursts of AI-generated political misinformation can propagate at extraordinary speed and scale.

But what does this misinformation look like? Consider some examples: in January 2024, voters in New Hampshire received automated phone calls that mimicked President Joe Biden's voice, urging them not to participate in the state's primary election.[59] The calls were later confirmed to be AI generated.[60] Around the same period, fake reports of an explosion near the Pentagon circulated online, accompanied by AI-generated images showing a large plume of smoke.[61] These images went viral within minutes and triggered an immediate market reaction. Between 10:04 a.m. and 10:10 a.m., both the Dow Jones Industrial Average and the S&P 500 fell sharply before fully recovering once the images were identified as fake.[62] In that six-minute interval, an estimated $500 billion in market value was temporarily erased.[63] Numerous social media accounts, including accounts linked to the Kremlin and Russian intelligence services, spread this fake news globally before being formally debunked by the Department of Defense and the Arlington County Fire Department.[64] Such misinformation also targets particular racial[65] and ethnic groups,[66] circulating false videos depicting Black women arguing with retail employees over public benefits, stealing from grocery stores, or boasting about receiving welfare assistance while unemployed;[67] it also includes AI-generated footage portraying Pope Leo XIV endorsing Captain Ibrahim Traore, a military leader who came to power in a 2022 coup.[68]

---

[53] *See e.g.,* Robert Chesney, Danielle K. Citron, *Deep Fakes: A Looming Challenge for Privacy, Democracy, and National Security*, 107 CALIF. L. REV 1753 (2019).
[54] *See* Soroush Vosoughi, Deb Roy and Sinan Aral, *The Spread of True and False News Online*, 359 SCIENCE 1146 (2018).
[55] *See* Chengcheng Shao et al., *The Spread of Low-Credibility Content by Social Bots*, 9 NATURE COMMUNICATIONS 4787 (2018).
[56] *See* Andrew Guess, et al., *Less Than You Think: Prevalence and Predictors of Fake News Dissemination on Facebook,* 5 SCIENCE ADVANCES (2019).
[57] *See* Alexander Romanishyn et al., *AI-Driven Disinformation: Policy Recommendations for Democratic Resilience,* FRONT ARTIL INTELL. (2025).
[58] *Id.*
[59] *See* Em Steck and Andrew Kaczynski, *Fake Joe Biden Robocall Urges New Hampshire Voters Not to Vote in Tuesday's Democratic Primary*, CNN (Jaunuary 22, 2024) https://edition.cnn.com/2024/01/22/politics/fake-joe-biden-robocall.
[60] *Id.*
[61] *See* Philip Marcelo, *Fact Focus: Fake Image of Pentagon Explosion Briefly Sends Jitters Through Stock Market*, AP (May 23, 2023), https://apnews.com/article/pentagon-explosion-misinformation-stock-market-ai-96f534c790872fde67012ee81b5ed6a4;

False political information of this kind is plainly detrimental to democratic self-government. It incites violence,[69] entrenches group stereotypes,[70] intensifies social polarization,[71] and undermines democratic commitments to openness and transparency. It also erodes trust in public institutions,[72] which has been cultivated over decades through a range of practices, including educational norms that

---

Donie O'Sullivan & Jon Passantino, *'Verified' Twitter Accounts Share Fake Image of 'Explosion' Near Pentagon, Causing Confusion*, CNN BUS. (May 23, 2023) https://www.cnn.com/2023/05/22/tech/twitter-fake-image-pentagon-explosion/index.html
[62] *Id. See* also Shannon Bond, Fake Viral Images of an Explosion at the Pentagon Were Probably Created by AI, NPR (May 22, 2023), https://www.npr.org/2023/05/22/1177590231/fake-viral-images-of-an-explosion-at-the-pentagon-were-probably-created-by-ai.
[63] *See* Ian Krietzberg, *S&P Sheds $500 Billion from Fake Pentagon Explosion*, THE STREET (May 22, 2023), https://www.thestreet.com/technology/s-p-sheds-500-billion-from-fake-pentagon-explosion.
[64] *See* Andrew Ross Sorkin et al., *An A.I.-Generated Spoof Rattles the Markets*, N.Y. TIMES (May 23, 2023), https://www.nytimes.com/2023/05/23/business/ai-picture-stock-market.html; Kierra Frazier, *Fake Image of Pentagon Explosion Goes Viral*, POLITICO (May 22, 2023), https://www.politico.com/news/2023/05/22/fake-image-pentagon-explosion-goes-viral-00098207
[65] *See* Kat Tenbarge, *Fake News YouTube Creators Target Black Celebrities with AI-Generated Misinformation,* NBC NEWS (Jan 30, 2024), https://www.nbcnews.com/tech/misinformation/ai-deepfake-fake-news-youtube-black-celebrities-rcna133368.
[66] *See* Alex Woodward, *Official White House Account Post Pixar-Style AI Meme Targeting Somalis After Claims of Fraud at Minnesota Daycares*, INDEPENDENT (January 2, 2026), https://www.the-independent.com/news/world/americas/us-politics/trump-somali-minnesota-daycare-fraud-ai-video-b2893624.html.
[67] *See* Irving Washington, Hagere Yilma, Joel Luther, *Fake AI-Generated Videos Perpetuate Stereotypes About SNAP Recipients, And New KFF Poll Looks at Belief in the False Claim that Undocumented Immigrants Are Eligible for ACA Coverage*, KFF (November 24, 2025) https://www.kff.org/health-information-trust/fake-ai-generated-videos-perpetuate-stereotypes-about-snap-recipients-and-new-kff-poll-looks-at-belief-in-the-false-claim-that-undocumented-immigrants-are-eligible-for-aca-coverage/.
[68] *See* Hannah Brockhaus, *Vatican Struggles Against Spread of "Deepfake" Images of Pope Leo XIV*, CATHOLIC NEWS AGENCY (Sep 25, 2025), https://www.catholicnewsagency.com/news/266765/vatican-struggles-against-spread-of-deepfake-images-of-pope-leo-xiv.
[69] *See* Gordon Corera & Brian Wheeler, *AI Could "Supercharge" Election Disinformation, US Tells the BBC*, BBC (February 14, 2024), https://www.bbc.com/news/world-68295845.
[70] *See* Gao et al., *Stereotypes in Artificial Intelligence-Generated Content: Impact on Content Choice*, APA PSYCNET (Sep 29, 2025).
[71] *See* Noah Lloyd, *New Research Reveals Algorithms' Hidden Political Power,* NORTHEASTERN GLOBAL NEWS (November 27, 2025), https://news.northeastern.edu/2025/11/27/social-media-political-polarization-research/.
[72] *Supra* Note 17.

prioritize reliable sources,[73] legal processes designed to elicit truthful information,[74] and administrative regimes that regulate, deter, and sanction the production of false or misleading claims.[75] It requires but little elaboration to demonstrate the necessity of regulating and deterring its transmission.[76]

## B. Regulations

To curb the dissemination of harmful AI-generated content online - content hosted by platforms and disseminated through social media, a significant portion of regulatory efforts has focused on platforms themselves.[77] And states, due to different political orientations and beliefs about whether false and harmful speech is best addressed through open circulation and counter-speech with minimal state interference, or whether misinformation is no longer self-correcting and instead requires affirmative government intervention, have accordingly adopted divergent laws that either require or prohibit content moderation on platforms.[78]

States that prohibit content moderation, such as Florida and Texas, have enacted anti-censorship laws. Florida's statute, for instance, S.B. 7072, applies to platforms with annual gross revenues exceeding $100 million or more than 100 million global monthly users and comprises five principal provisions.[79] First, it prohibits platforms from "willfully deplatform[ing]" a candidate for office,[80] with "deplatform"

---

[73] E.g., university librarians instructing students to prioritize ".gov" websites as among the most reliable sources on the web. *See Evaluating Your Sources*, MENLO COLLEGE BOWMAN LIBRARY (last updated Jan.2, 2026), https://library.menlo.edu/c.php?g=1123648&p=8195788.
[74] *See e.g.,* Tehan v. United States ex rel. Shott, 382 U.S. 406, 416 (1966) (one basic purpose of a trial is the determination of truth.")
[75] *See* Rebecca S. Eisenberg, *The Role of the FDA in Innovation Policy*, 13 MICH. TELECOMM. & TECH. L. REV. 345, 367 (2007) (noting the FDA's "modern function of getting [private] firms to conduct rigorous clinical trials of drugs"); Rachel E. Sachs, *Administering Health Innovation*, 39 CARDOZO L. REV. 1991, 1999-2000 (2018) (noting similar NIH requirements). *See also* 40 C.F.R. § 1066 (2023) (setting forth procedures by which auto makers must conduct emissions testing and requiring submission of test results to the EPA).
[76] Indeed, the World Economic Forum's *Global Risks Report* listed AI-powered misinformation and disinformation as the most severe threat to the world in the next two years. Even in the long term, this report rates AI-powered misinformation and "adverse outcomes of AI technologies" as the fifth- and sixth- greatest risks, respectively. The report reached this conclusion based on the input of 1,490 experts across academia, business, government, the international community and civil society. *See* David Inserra, *Artificial Intelligence Regulation Threatens Free Expression*, CATO INSTITUTE (July 16, 2024), https://www.cato.org/briefing-paper/artificial-intelligence-regulation-threatens-free-expression#new-technology-same-moral-panic.
[77] Other regulatory interventions exist, but they are largely confined to private law issues, particularly copyright and product liability, such as whether training data may be lawfully scraped and whether model outputs infringe protected works, and if developers should bear responsibility for harmful outputs.
[78] For example, progressive states such as California and New York would require content moderation. *See, e.g.,* sec. 2, § 1798.99.31, 2022 Cal. Stat. at 4920–22; Act of June 6, 2022, ch. 204, § 1, 2022 N.Y. Laws 1176, 1177 (codified at N.Y. GEN. BUS. LAW § 394-ccc (2023)); S. 287, 2023-24 Leg., Reg. Sess. (Cal. 2023). Conservative states such as Texas and Wisconsin would prohibit it. See, e.g., TEX. CIV. PRAC. & REM. CODE ANN. § 143A.002(a) (Westlaw through legislation effective July 1, 2023, of the 2023 Reg. Sess. of the 88th Leg.); H.R. 441, 134th Gen. Assemb., Reg. Sess. (Ohio 2021); Assemb. 530, 2021–22 Leg., Reg. Sess. (Wis. 2021) (proposing a ban on censoring "official pages, accounts, profiles, or handles relating to a candidate's campaign or an elected official's office"); Assemb. 589, 2021–22 Leg., Reg. Sess. (Wis. 2021) (proposing a ban on censoring journalistic enterprises based on the content of their publications or broadcasts).
[79] FLA. STAT. § 501.2041(1)(g)(4)(a)–(b) (2022).
[80] FLA. STAT. § 106.072(2).

defined as deleting or banning a user for more than fourteen days.[81] Second, it prohibits the use of "shadow banning" and "post-prioritization" for content posted by or about a political candidate.[82] Shadow banning and post-prioritization refer to content-moderation practices through which platforms prioritize some posts over others.[83] Third, the statute bars "censor[ship], deplatform[ing], or shadow ban[ning]" of a "journalistic enterprise based on the content of its publication or broadcast," unless that content is obscene.[84] Fourth, platforms must apply censorship, deplatforming, and shadow-banning standards in a consistent manner across users.[85] Fifth, platforms are required to "categorize" post-prioritization and shadow-banning algorithms and to offer users the opportunity to opt out of those algorithms at least once per year.[86]

Texas has enacted similar legislation. H.B. 20 prohibits "social media platforms" from "censor[ing]" a user's expression, or a user's ability to receive the expression of another person, on the basis of viewpoint or geographic location.[87] "Social media platforms" include all "internet websites or applications" that are "open to the public, allow a user to create an account, and enable users to communicate with other users for the primary purpose of posting information, comments, messages, or images."[88] This prohibition also applies where the viewpoint is expressed outside the platform, meaning that platforms may not remove users or content based on speech occurring elsewhere.[89] Both states justify these restrictions by analogizing social media platforms to "common carriers,"[90] whose obligations require them to offer services to nearly all comers, much like inns, railroads, and telecommunications services.

States that require content moderation include New York,[91] and California.[92] Take California as an example. To compel platforms to moderate more aggressively, the state enacted the California Age-Appropriate Design Code Act, which imposes stringent obligations on online services "that children are likely to access."[93] The Act requires platforms to enforce their content policies where those policies apply to California users.⁹⁴ It further prohibits platforms from using a child's personal information in any manner the business "has reason to know is materially detrimental to the physical health, mental health, or well-being of a child."⁹⁵ Although the statute does not define these terms with precision, it establishes a state agency empowered to develop "best practices" that may supply interpretive guidance.⁹⁶ This Act also restricts the use of automated systems to process children's personal information. Platforms may not deploy such systems unless they can demonstrate that appropriate safeguards are in place to protect children and can further show either that the automated processing is necessary to provide the product or service, or that there is a "compelling reason" for its use.⁹⁷

---

[81] *Id.* § 501.2041(1)(c).
[82] *Id.* § 501.2041(2)(h).
[83] *Id.* § 501.2041(1)(e)–(f).
[84] *Id.* § 501.2041(2)(j).
[85] FLA. STAT. § 501.2041(2)(b).
[86] *Id.* § 501.2041(2)(f)–(g).
[87] Sec. 2, § 143A.002(a).
[88] Act of Sept. 9, 2021, ch. 3, sec. 2, § 120.001(1) (codified at TEX. BUS. & COM. CODE ANN. § 120.001(1) (2021)).
[89] Sec. 2, § 143A.002(b).
[90] *See* Act of May 24, 2021, ch. 32, § 1(6), 2021 FLA. LAWS 503, 505 ("Social media platforms . . .should be treated similarly to common carriers.").Sec. 1(3)–(4), 2021 TEX. GEN. LAWS 2d Sess. at 3904 (finding social media platforms to be common carriers).
[91] *See* Act of June 6, 2022, ch. 204, 2022 N.Y. Laws 1176 (codified at N.Y. GEN. BUS. LAW § 394-ccc (2023)) (prescribing social media networks institute process for users to report "hateful conduct" and accessible policy for responding to such reports).
[92] *See* California Age-Appropriate Design Code Act, ch. 320, sec. 2, § 1798.99.31, 2022 Cal.
Stat. 4916, 4920–22 (effective July 1, 2024) (to be codified at CAL. CIV. CODE § 1798.99.31) (requiring platforms "likely to be accessed by children" to enforce their promulgated policies, terms of service, and community standards)
[93] Sec. 2, § 1798.99.29(a), 2022 Cal. Stat. at 4919 (codified at CAL. CIV. CODE § 1798.99.29(a) (2023)).

States also vary in their platform disclosure requirements. Florida's S.B. 7072 imposes detailed notice obligations, requiring platforms to provide written notice —"includ[ing] a thorough rationale explaining the reason that the social media platform" moderated a user's content and "a precise and thorough explanation of how the social media platform became aware of the censored content or material"— within seven days of censoring, shadow banning, or deplatforming a user.⁹⁸ Texas, by contrast, adopts a more general disclosure regime. Its H.B. 20 requires platforms to publish biannual transparency reports containing aggregate data on the number of user complaints received regarding unlawful or policy-violating content and the platforms' responses to such complaints.⁹⁹

The California Age-Appropriate Design Code Act likewise includes disclosure obligations, though of a different kind. Covered platforms must conduct a "Data Protection Impact Assessment" before launching any new feature, must review that assessment at least twice per year, and must provide it to the state upon request.¹⁰⁰ The assessment must identify the purpose of the new feature and evaluate eight categories of potential harms to children, including exposure to potentially harmful content, contacts, conduct, algorithms, and targeted advertising.¹⁰¹ Notably, the statute does not define what constitutes "harmful" content, algorithms, or advertisements, leaving those terms largely open-ended.

---

⁹⁴ Sec. 2, § 1798.99.31(a)(9), 2022 Cal. Stat. at 4921 (effective July 1, 2024) (to be codified at CAL. CIV. CODE § 1798.99.31(a)(9)) (requiring that businesses "[e]nforce published terms, policies, and community standards established by the business").
⁹⁵ Sec. 2, § 1798.99.31(b)(1), 2022 Cal. Stat. at 4921.
⁹⁶ Sec. 2, § 1798.99.32(d), 2022 Cal. Stat. at 4922 (codified at CAL. CIV. CODE § 1798.99.32(d)).
⁹⁷ Sec. 2, § 1798.99.31(b)(2)(A)–(B), 2022 Cal. Stat. at 4921 (effective July 1, 2024) (to be codified at CAL. CIV. CODE § 1798.99.31(b)(2)(A)–(B)) (banning profiling of a child except in limited circumstances); see also sec. 2, § 1798.99.30(b)(6), 2022 Cal. Stat. at 4919 (codified at CAL. CIV. CODE § 1798.99.30(b)(6)) (defining "profiling" as any "automated processing of personal information that uses personal information to evaluate certain aspects relating to a natural person").
⁹⁸ See Act of May 24, 2021, ch. 32, § 4, 2021 Fla. Laws 503, 513–14 (codified as amended at FLA. STAT. § 501.2041(2)–(3) (2022)).
⁹⁹ Sec. 2, § 120.053, 2021 Tex. Gen. Laws 2d Sess. at 3906–07. It also requires platforms to publicly disclose information about how they curate content and use algorithms. Sec. 2, §§ 120.051, .053, 2021 Tex. Gen. Laws 2d Sess. at 3905–07. A different Texas law requires platforms to disclose "an overview of" how the platform uses algorithms for content moderation. See Securing Children Online Through Parental Empowerment Act, ch. 795, § 2.01, 2023 Tex. Sess. Law Serv. (West) (effective Sept. 1, 2024) (codified at TEX. BUS. & COM. CODE ANN. § 509.056 (2021)).
¹⁰⁰ California Age-Appropriate Design Code Act, ch. 320, sec. 2, § 1798.99.31(a)(1)(A)–(B), (a)(3), 2022 Cal. Stat. 4916, 4920 (effective July 1, 2024) (to be codified at CAL. CIV. CODE § 1798.99.31(a)(1)(A)–(B), (a)(3)).
¹⁰¹ Sec. 2, § 1798.99.31(a)(1)(B)(i)–(viii), 2022 Cal. Stat. at 4921.

Other state[102] and federal statutes[103] targeting the creation and dissemination of harmful content, whether AI-generated or not, likewise exist, though they largely follow the same regulatory framework and are omitted here to avoid redundancy.

## PART II. First Amendment Protects AI-Generated Harmful Content

This Part argues that, despite the serious and well-documented harms associated with AI-generated sexually explicit content and political misinformation, such expression remains protected by the First Amendment under existing doctrine. The analysis includes four steps: first, it addresses the threshold question whether the Free Speech Clause applies to AI-generated content at all and answers that question in the affirmative. It then examines, in steps two through four, whether any of the established categorical exceptions to First Amendment protection, including obscenity, hate speech, and fighting words, apply to the content at issue, concluding that none does in a meaningful way and that the expression remains constitutionally protected.

---

[102] Minnesota enacted statutes banning or requiring disclosure of AI-generated deepfakes in election advertising. Virginia broadened its nonconsensual pornography statute to cover "falsely created" or computer-generated intimate images. Many states have also begun regulating interaction with LLM-based chatbots, requiring disclosure that users are speaking with a machine or imposing safeguards in sensitive contexts such as health and therapy. *See* Minn. Stat. § 211B.075; Cal. Elec. Code § 20010; Va. Code. Ann. §18.2-386.2 (2020) (criminalizing falsely created pornographic images); H.B. 2678, 2019 Leg., Reg. Sess. (Va. 2019) (providing the amended language to Virginia's original law criminalizing the malicious distribution of pornographic images without the subject's consent). *See also* Robert Volker & Henry Ajder, Analyzing the Commoditization of Deepfakes, 2020 N.Y.U. J. LEGIS. & PUB. POL'Y QUORUM 22, 27 (crediting Virginia as the first state to criminalize "nonconsensual, 'falsely created,' explicit images and videos [as] . . . a Class 1 misdemeanor"); California Bot Disclosure Law (CAL. BUS. & PROF. CODE §17940-17942) (if a bot is used to interact with a person with the intent to mislead into thinking it's human, the deployer must "clearly and conspicuously" disclose that it is a bot); Utah, Artificial Intelligence Policy Act (requires businesses using generative AI to disclose when interacting with users, prohibiting "blaming" AI for harmful statements); Maine, AI Chatbot Disclosure Act (when a business uses chatbot/AI that might mislead a "reasonable consumer" into thinking it is human, must disclose that it is AI); Colorado, AI-consumer interaction disclosure (proposed) (requires AI systems that interact with consumers to carry a disclosure that the consumer is interacting with AI); SB 640 (would require any business using a chatbot to disclose to consumers that they are not interacting with a human when the bot may mistead); Hawaii HB 639/SB 640 (among proposed bills requiring disclosure when chatbot used in consumer interactions).

[103] For Federal statutes, consider Section 230 of the Communications Decency Act, which shields platforms from liability for third-party content while allowing them, in good faith, to restrict material they deem objectionable. Congress also enacted the Take it Down Act of 2025, which was the first federal statute to regulate both authentic and AI-generated intimate images, to make it unlawful for any person "to use an interactive computer service to knowingly publish an intimate visual depiction of an identifiable individual" under specified circumstances. The Act also prohibits threats to publish nonconsensual intimate imagery (NCII) and imposes criminal penalties for violations. Platforms qualified as "covered services" must comply with a mandatory notice-and-takedown process requiring them to accept removal requests from victims, respond within 48 hours, and make reasonable efforts to eliminate duplicates. *See* Richard B. Newman, Hinch Newman, *Federal Take it Down Act Targeting Revenge-Porn Becomes Law*, NATIONAL LAW REVIEW (May 26, 2025), https://natlawreview.com/article/federal-take-it-down-act-targeting-revenge-porn-becomes-law.

## A. Does First Amendment Apply to AI-Generated Content?

The question whether the First Amendment applies to AI-generated content has attracted significant debate in recent years.[104] This Section argues that the First Amendment does apply to AI-generated content for three reasons:

First, AI-generated content such as posts, tweets, images, artwork, reviews, and other expressive media qualifies as protected speech insofar as it bears on democratic self-government.[105] Such content is unlike that generated by the systems that merely assist users in performing instrumental tasks, such as car alarms, GPS routing, or system notifications, which do not convey ideas.[106] The examples just described contribute to the production and dissemination of ideas and thus participate in the marketplace of ideas the First Amendment is designed to sustain; that they may be misleading, offensive, or otherwise unsavory does not alter their constitutional character.

Second, listeners have a constitutionally cognizable interest in receiving content that informs their ideas.[107] Empirical literature and the World Economic Forum have documented that AI-generated content can improve medical diagnosis,[108] expand access to mental-health support,[109] and deliver informational and advisory services to underserved communities.[110] To the extent such content advances the audience's ability to receive information and test ideas, it serves the listener interests the First Amendment exists to protect and therefore falls within its scope.[111] The nonhuman character of the speaker is beside the point. The Court has long extended First Amendment protection to corporate speakers precisely because their expression contributes to public discourse.[112] Earlier communicative technologies, from the printing press to broadcast media, likewise initially met skepticism before receiving constitutional protection once their expressive value became apparent.[113] AI fits comfortably within this historical pattern.

---

[104] *See e.g.,* Stuart Minor Benjamin, *Algorithms and Speech*, 161 U.PA. L. REV 1445 (2013) (arguing that algorithmic outputs can be speech because the First Amendment has never required consciousness); Karl M. Manheim & Jeffrey Atik, *AI Outputs and the Limited Reach of the First Amendment*, 63 WASHBURN L.J. 159 (2024) (skeptical that AI outputs are not expressive acts deserving constitutional protection); Amanda Peters, *Why an AI Editor Does Not Serve First Amendment Values,* 95 U. COLO.L.REV. 307 (2024) (arguing that even if AI looks like an editor, it does not advance the normative commitments underlying First Amendment Protection). For popular media discussing this issue, *see e.g., Artificial Intelligence, Free Speech, and the First Amendment,* FOUNDATION FOR INDIVIDUAL RIGHTS AND EXPRESSION, https://www.thefire.org/research-learn/artificial-intelligence-free-speech-and-first-amendment (last updated Jan. 3, 2026); Ashkhen Kazaryan, *AI and 1A: is Artificial Intelligence Protected by the First Amendment*, FREEDOM FORUM (May 22, 2025), https://www.freedomforum.org/artificial-intelligence-first-amendment/; Scott Bomboy, *Lawsuit Analyzes First Amendment Protection for AI Chatbots in Civil Case*, NATIONAL CONSTITUTION CENTER (May 7, 2025), https://constitutioncenter.org/blog/lawsuit-analyzes-first-amendment-protection-for-ai-chatbots-in-civil-case.
[105] *See* Tim Wu, *Machine Speech*, 161 U. PA. L. REV. 1495, 1524 (2013).
[106] *Id* at 1525.
[107] *See* Eugene Volokh, Mark A. Lemley & Peter Henderson, *Freedom of Speech and AI Output,* 3 J. FREE SPEECH L. 651 (2023); *See also AI and the First Amendment: A Q&A with Jack Balkin*, YALE LAW SCHOOL (January 29, 2024), https://law.yale.edu/yls-today/news/ai-and-first-amendment-qa-jack-balkin.
[108] *See* Jinseo Jeong, *Reducing the Workload of Medical Diagnosis Through Artificial Intelligence: A Narrative Review*, 104 MEDICINE (BALTIMORE) No. 6, e41470 (2025).
[109] *See* Hailey Fowler and John Lester, *How AI Could Expand and Improve Access to Mental Health Treatment*, WORLD ECONOMIC FORUM (Nov 1, 2024), https://www.weforum.org/stories/2024/10/how-ai-could-expand-and-improve-access-to-mental-health-treatment/.
[110] *See* Toluwani Aliu, *How AI is Powering Grassroots Solutions for Underserved Communities,* WORLD ECONOMIC FORUM (Sep 2, 2025), https://www.weforum.org/stories/2025/09/how-ai-is-powering-grassroots-solutions-for-community-challenges/.
[111] *Supra* Note 107.

Third, chatbots, though often described as autonomous systems driven solely by statistical computation across latent spaces and attention mechanisms,[114] respond to human prompts and operate within systems designed, trained, and deployed by human actors.[115] Humans are an indispensable part of this communicative loop. User disclosures on platforms become training data;[116] models generate outputs that users consume, interpret, and act upon;[117] and those responses, in turn, produce further disclosures. Humans are thus both the sources and the recipients of the data that sustains this cycle.[118] To deny First Amendment protection in this context is to excise human agency from an emergent communicative environment.[119]

## B. Is there an Exception for Nonconsensual Pornography?

Turning to the question whether AI-generated sexually explicit images fall within any recognized exception to First Amendment protection, the answer remains largely negative, for three reasons: first, although such content may be unsavory, it is not obscene within the meaning of obscenity doctrine; second, the compelling-interest rationale that sustains nonconsensual-pornography statutes does not readily transfer to fictional sexual content generated by AI. Finally, there is no historical tradition supporting the creation of a new categorical exception for machine-generated sexual images. Under *United States v. Stevens* and *Brown v. Entertainment Merchants Association*, the absence of such a tradition sharply constrains judicial authority to expand the scope of unprotected speech.[120]

---

[112] *See* First Nat'l Bank of Boston v. Bellotti, 435 U.S. 765 (1978).
[113] *See generally* Ronald K. L. Collins & David M. Skover, ROBOTICA: SPEECH RIGHTS AND ARTIFICIAL INTELLIGENCE (John Berger ed., Cambridge Univ. Press 2018).
[114] *See* Mackenzie Austin & Max Levy, *Speech Certainty: Algorithmic Speech and the Limits of the First Amendment*, 77 STAN. L REV. 1 (2025).
[115] *See* Benjamin, *Algorithms and Speech*, supra note 104, at 1479 ("The fact that an algorithm is involved doesn't mean that a machine is doing the talking."); Toni M. Massaro & Helen Norton, *Siri-ously? Free Speech Rights and Artificial Intelligence*, 110 NW. U. L. REV. 1169 (2016); *See also* Toni M. Massaro, Helen Norton & Margot E. Kaminski, *Siri-ously 2.0: What Artificial Intelligence Reveals About the First Amendment*, 101 MINN. L. REV. 2481 (2017).
[116] This process is called web scrapping, which involves fetching a webpage and extracting useful information from it for data analysis. On a high-level, it involves the steps such as: 1) sending a request to the target website server to fetch the webpage's content, 2) downloading the HTML content, 3) parsing the HTMP, 4) Extracting relevant data, and 5) storing the data in various formats such as CSV, JSON, or any other data structures for further processing.
[117] Such as building relationships with language models, listening to their advice, use it as life counsels. *See e.g.,* Lance Eliot, *Using Generative AI to Perform Life Reviews at Any Stage of Life*, FORBES (Sep 3, 2024), https://www.forbes.com/sites/lanceeliot/2024/09/03/using-generative-ai-to-perform-life-reviews-at-any-stage-of-life/.
[118] For an overview of how individuals both serve as authors of data and being used as recipients of results from data analysis, *see generally* SHOSHANA ZUBOFF, *The Age of Surveillance Capitalism: The Fight for a Human Future at the New Frontier of Power* (2019).
[119] It only falls outside the scope of First Amendment if a system were one day to become independent enough that no human can be said to be involved in the production cycle. Benjamin, *Algorithms and Speech*, supra note 104 at 1481-82.
[120] *See* Brown v. Ent. Merchs. Ass'n, 564 U.S. 792, 795 (2011) (Court refusing to recognize an exception for violent video games sold to minors, because it found no tradition of restrictions on violent speech to minors; United States v. Stevens,

The Court has never recognized a categorical exclusion of sexually explicit material from First Amendment protection.[121] Pornographic expression loses constitutional protection only if it qualifies as obscene under *Miller v. California*,[122] which requires that the material, taken as a whole, (1) appeal to the prurient interest, (2) depict sexual conduct in a patently offensive manner, and (3) lack serious literary, artistic, political, or scientific value.[123] The prurient interest inquiry focuses on the dominant character of the work as a whole and asks whether it is designed to appeal to a shameful or morbid interest in sex, as judged by contemporary community standards, rather than on the subjective intent of individual creators.[124] At present, many AI-generated sexual images are produced for purposes such as experimentation, parody, aesthetic play, model probing, or sheer novelty,[125] rather than for sexual gratification. The same prompt may yield outputs that are incoherent, uncanny, satirical, or emotionally flat.[126] In this context, the dominant character of the material, taken as a whole, is not an appeal to prurient interest but an exercise in technical exploration or expressive bricolage.[127] Such content therefore fails the prurient-interest prong of the *Miller* test.

Even assuming arguendo that obscenity doctrine could apply, the rationales courts have relied upon to uphold regulations banning sexually explicit material do not readily extend to AI-generated content. As of 2025, all fifty states and the District of Columbia prohibit nonconsensual pornography.[128]

---

559 U.S. 460 (2010) The government urged the Court to recognize a categorical exception to First Amendment protection for depictions of animal cruelty, arguing that all new "claim[s] of categorical exclusion should be considered under a simple balancing test . . . balancing the value of the speech against its societal costs." *Id* at 470. The Court declined to recognize the government's proposed exception because, though "the prohibition of animal cruelty itself has a long history in American law . . . [there is no] similar tradition excluding depictions of animal cruelty from 'the freedom of speech' codified in the First Amendment." *Id.* at 469

[121] *Compare* New York v. Ferber, 458 U.S. 747, 764 (1982) (restricting pornography that visually depicts children below the age of majority performing sexual acts) *with* Ashcroft v. Free Speech Coal., 535 U.S. 234(2002) (virtual child pornography is protected speech under the First Amendment).

[122] 413 U.S. 15, 23-24 (1973) (holding that "obscene" material not protected by first amendment). See also *Ashcroft* at 240 (2002) (noting that, as "a general rule, pornography can be banned only if obscene"); *See, also*, Paris Adult Theater I v. Slaton, 413 U.S. 49, 56 (1973) (same); Roth v. United States, 354 U.S. 476, 481 (1957) (federal statute criminalizing "obscenity" was not violative of first amendment); Chaplinsky v. New Hampshire, 315 U.S. 568, 571-72 (1942) (certain classes of speech such as "the lewd and the obscene" not afforded first amendment protection).

[123] *Id.*

[124] *Id Roth* at 489.

[125] *See* Asher Flynn et al., *Sexualized Deepfake Abuse: Perpetrator and Victim Perspectives on the Motivations and Forms of Non-Consensually Created and Shared Sexualized Deepfake Imagery*, J. INTERPERSONAL VIOLENCE (2025) (identifying "curiosity" and "experimentation" as one of the self-reported motivations, with perpetrators sometimes minimizing harms and framing their behavior as "a joke" or exploration of the tech).

[126] This is because of the nature of the generative models, that language models produce a probability distribution over a set of words and then determining the next word by *sampling* from this distribution. So, determining the next word is stochastic.

[127] *See New Research on How People Are Interacting with GenAI Sexual Content*, SNAP VALUES (November 19, 2024), https://values.snap.com/news/new-gen-ai-research (a consumer survey showing that a substantial share of teens and young adults have *seen* or *encountered* AI-generated sexual images, without typing that exclusively to harmful intent. For instance, some teen girls in particular are feeling "left out" if they are not featured in AI-manipulated sexual images that their peers are inappropriately creating.

[128] *See Nonconsensual Distribution of Intimate Images*, CYBER CIVIL RIGHTS INITIATIVES, https://cybercivilrights.org/nonconsensual-distribution-of-intimate-images/#:~:text=*%2050%20States%20+%20DC%20+%20Two,Massachusetts%20District%20of%20Columbia.%20Back%20to%20Map.(last updated Jan. 3, 2026).

Although many of these statutes have faced First Amendment challenges, courts have largely sustained them under strict scrutiny based on a compelling interest in preventing concrete and individualized harms caused by knowing disclosure of intimate images without consent.[129] These harms include the destabilization of personal identity, severe emotional distress,[130] corrosion of self-worth, social isolation,[131] the alteration of victims' perceived standing within their communities, and a broader signal of societal indifference to emotional safety and bodily autonomy.[132]

Those interests, however, do not readily translate to AI-generated sexual content. Where no real person is depicted, there is no invasion of privacy, no violation of bodily autonomy, and none of the harms that motivate the nonconsensual-pornography cases. Even when an AI-generated image is mapped onto a real person, substantial doctrinal obstacles remain: such images do not purport to assert that the depicted individual actually engaged in the sexual conduct shown, nor do they involve the dissemination of private intimate material obtained from the subject. Rather, the data used to generate these images are typically drawn from publicly available sources, such as social media accounts whose content the owners themselves made public, rather than confidential personal exchanges of the kind that nonconsensual-pornography statutes were designed to address. Absent a concrete invasion of privacy or an individualized injury, the compelling-interest justification that sustains these statutes under strict scrutiny fails.

## C. Does the First Amendment Protect Hate-Based Political Misinformation?

AI-generated political misinformation, as described above, does not become regulable merely because it attacks or debases groups on the basis of race, ethnicity, religion, gender, or sexual orientation. American constitutional law has never recognized hate speech as a freestanding category of unprotected expression, and that doctrinal position remains unchanged.[133]

As a matter of law, regulations targeting hateful expression are treated as content-based restrictions subject to the most exacting scrutiny, a standard that has made it extraordinarily difficult for state or local governments to regulate such speech. The few doctrinal moments that once appeared to signal greater judicial tolerance for such regulation did not last. In *Beauharnais v. Illinois*, the Court upheld a "group libel" statute that criminalized publications portraying racial or religious groups as depraved or criminal.[134] And in *Snyder v. Phelps*, the Court suggested that speech inflicting severe emotional distress might be regulatable when it does not address a matter of public concern.[135] Neither case, however, altered First Amendment doctrine in any substantive way. *Beauharnais* is now regarded as an historical anomaly,[136] and *Snyder* gestured at, but did not establish, an intentional-infliction-of-emotional-distress exception for Free Speech. As a doctrinal matter, hate speech remains protected. The normative literature advocating for a hate-speech exception is richer,[137] but it has not moved the doctrine. Jeremy Waldron argues that such speech "disfigure[s] our social environment by [communicating] that . . . members of [certain social] group[s] are not worthy of equal citizenship."[138] It creates collective identities of superiority and subordination and shapes, in corrosive ways, the character of the society through which it circulates.[139] Mari Matsuda likewise emphasizes both persuasive and constitutive harms: victims leave jobs, schools, neighborhoods, and public spaces;[140] ideas of inferiority are replanted even in those who consciously resist them;[141] repetition "interfer[e]s with our perception and interaction" with others and erodes the sense of common humanity.[142] But such arguments are nevertheless theoretical. Without a historical tradition of regulating this category of speech, bound by *Stevens* and *Brown*, there's not much the Court may do. It may just be so that AI-generated hate speech targeting specific groups remains protected under strict scrutiny.[143]

---

[129] People v. Austin, 155 N.E.3d 439, 454–60 (Ill. 2019) (identifying a compelling interest in preventing "serious emotional distress" and reputational harm caused by knowing, nonconsensual disclosure of private sexual images), State v. VanBuren, 214 A.3d 791, 805–15 (Vt. 2019) (sustaining Vermont's statute under strict scrutiny, stressing individualized harms, the absence of public concern, and the statute's focus on intentional disclosure without consent), State v. Casillas, 458 P.3d 1257, 1266–73 (N.M. Ct. App. 2019) (applying strict scrutiny and finds a compelling interest in protecting individuals from the uniquely severe harms of nonconsensual disclosure of intimate imagery).
[130] *VanBuren*, 214 A.3d at 810.
[131] Austin at 461 (Ill. 2019).
[132] Casillas at 641-642 (Minn. 2020).
[133] *See* Eugene Volokh, *No, There's No "hate Speech" exception to the First Amendment*, WASH. POST (May 7, 2015, 6:02 PM), https://www.washingtonpost.com/news/volokh-conspiracy/wp/2015/05/07/no-theres-no-hate-speech-exception-to-the-first-amendment/; Eugene Volokh, *Burning to Say Something*, WALL STREET JOURNAL (April 9, 2003).

---

[134] 343 U.S. 250, 251 (1952).
[135] 562 U.S. 443 (2011). The issue here is that whether the father of a fallen solider could sue on state tort claim against members of the Westboro Baptist Church, who had protested near the soldier's funeral. *Id* at 448-50. The Court held that protesters could not be punished because their speech was on a matter of public concern. *Id* at 451-56. But the Court didn't decide that the speech cannot be proscribed if the issue was *not* on a matter of public concern. *Id* at 451.
[136] *See, e.g.,* Am. Booksellers Ass'n v. Hudnut III, 771 F.2d 323, 331 n.3 (7th Cir. 1985), summarily aff'd, 475 U.S. 1001 (1986) (noting the erosion of Beauharnais).
[137] *See, e.g.*, RONALD DWORKIN, *The Coming Battles Over Free Speech*, N.Y. Rev. Books, (1992) (discussing arguments that defend First Amendment rights via Justice Brennan and the landmark Sullivan case); Robert C. Post, *Racist Speech, Democracy, and the First Amendment*, 32 WM. & MARY L. REV. 267, 322 (1991) (same); Marcus Schulzke, *The Social Benefits of Protecting hate Speech and exposing Sources of Prejudice*, 22 RES PUBLICA 225, 225 (2016) (reasoning that hate speech may expose people's harmful attitudes); Calvert Magruder, *Mental and emotional Disturbance in the Law of Torts*, 49 HARV. L. REV. 1033, 1053 (1936) (arguing that hate speech is not low value speech, because it provides a safety valve through which irascible tempers might legally blow off steam). *See contra.* John C. Knechtle, *When to Regulate hate Speech*, 110 PENN ST. L. REV. 539, 543 (2006) ("[A]t a minimum, hate speech that threatens unlawful harm or incites to violence may be proscribed."); Andrew Reid, *Does Regulating Speech Undermine Democratic Legitimacy? A Cautious 'No'*, 26 RES PUBLICA 181, 181 (2020) ("[I]n some cases the harmful effects of hateful speech on the democratic process outweigh those of restriction."); Alexander Tsesis, *hate in Cyberspace: Regulating hate Speech on the Internet*, 38 SAN DIEGO L. REV. 817, 820 (2001) (stating that hate speech on the internet "should be prohibited"); Scott J. Catlin, *A Proposal for Regulating hate Speech in the United States: Balancing Rights under the International Covenant on Civil and Political Rights*, Note, 69 NOTRE DAME L. REV. 771, 771 (1994) (seeking an "approach to regulating hate speech").
[138] JEREMY WALDRON, *The Harm in Hate Speech* (Harvard Univ. Press 2012).
[139] David W. Seitz & Amanda Berardi Tennant, *Constitutive Rhetoric in the Age of Neoliberalism*, in Rhetoric in Neoliberalism 109 (Kim Hong Nguyen ed., 2017); Waldron, *Id* at 93. Waldron also recognizes that hate speech creates a "social environment" where "leaves [cannot] be led, … children [cannot] be brought up, [and] … hopes [cannot] be maintained." *Id* at 33.
[140] Mari J. Matsuda, Public Response to Racist Speech: Considering the Victim's Story 87 MICH. L. REV 2320, 2337 (1993).
[141] *Id* at 2339 ("plants [racial inferiority] in our minds as an idea that may hold some truth… no matter how much both victims and well-meaning dominant group members resist it.")
[142] *Id* (it "distances… dominant-group members from the victims, making it harder to achieve a sense of common humanity." And also, through the ideas hate speech communicate are "improbable and abhorrent… [they] interfere with our perception and interactions" [with others because they are] presented repeatedly.")
[143] Indeed, changing this situation may be a very daunting task, for the United States is one of the only developed countries that has no known history of hate speech law. The European Union, for example, prohibits pictures and writings that

## D. What of Fighting Words?

What about those words that are generated by AI, are particularly offensive and directed at specific individuals? Consider the following scenario. A state agency deploys an AI system at a political rally organized by an opposing party. Equipped with facial recognition technology, geofencing, and similar identification tools, the system identifies named individuals physically present in the crowd and then generates and broadcasts hyper-personalized, degrading audio messages directed exclusively at those individuals. The messages include calling them "worthless parasites" who "do not belong in this country," mocking their appearance as a "disgusting freak," targeting their family by declaring it "a disgrace known for the wrong reasons," and condemning their political affiliation by asserting that "people like you ruin this nation and should be driven out of public life." Would such AI-generated speech remain protected under the Free Speech Clause?

Most likely yes, under current doctrine. Although such speech would appear, in doctrinal terms, to satisfy the classic formulation of fighting words insofar as it "tend[s] to incite an immediate breach of the peace" by provoking a face-to-face confrontation[144] and consists of "personally abusive epithets" directed to the person of the hearer and reasonably understood as direct personal insults likely to provoke a violent reaction,[145] the Supreme Court has not relied on the fighting-words doctrine as a sustained basis for conviction since it first announced the category in *Chaplinsky v. New Hampshire* in 1942.[146]

---

incite "hatred directed against a group… defined by reference to race, colour, religion, descent, or national or ethnic origin." Countries in Asia, Oceania, and North America have similar laws. *See* Council Framework Decision 2008/913/JHA of Nov. 28, 2008, On Combating Certain Forms and Expressions of Racism and Xenophobia by Means of Criminal Law, 2008 O.J. (L 328) 55, 56; *see also* Communication From the Commission to the European Parliament and the Council: A More Inclusive and Protective Europe: extending the List of EU Crimes to hate Speech and hate Crime, COM (2021) 777 final (Dec. 9, 2021). Koji Higashikawa, *Japan's hate Speech Laws: Translations of the Osaka City Ordinance and the National Act to Curb hate Speech in Japan*, 19 ASIAN-PACIFIC L. & POL'Y J. 1, 4 (2017); *see also* Junko Kotani, *Proceed with Caution: hate Speech Regulation in Japan*, 45 HASTINGS CONST. L.Q. 603, 604 (2018) (explaining that "[t]he law narrowly defines hate speech and declares it inappropriate and impermissible, but it does not criminalize or illegalize such speech, nor does it have a built-in system through which the law can be enforced"). *See* Katharine Gelber & Luke McNamara, *The effects of Civil hate Speech Laws: Lessons from Australia*, 49 L. & SOC'Y REV. 631, 634 (2015) (cataloging a chronology of Australian hate speech laws). Canada prescribes criminal punishments for "everyone who, by communicating statements, . . . willfully promotes hatred against any identifiable group," Criminal Code, R.S.C. 1985, c C-46 319(2), (2.1) (Can.), or "willfully promotes antisemitism by condoning, denying or downplaying the Holocaust."

[144] Chaplinsky v. New Hampshire, 315 U.S. 568, 572 (1942) (the Court upheld the conviction of a man who had called a city official a "God damned racketeer" and a "damned Fascist," because the phrases were especially "likely to provoke the average person to retaliation" or to persuade individuals to "cause a breach of the peace.")

[145] Cohen v. California, 403 U.S. 15, 20 (1971) (a man was arrested for wearing a jacket that said "Fuck the Draft" into a courthouse. The Supreme Court reversed the conviction, finding that the fighting-words exception did not apply. The Court noted that the message was not "directed at the person of the hearer" and was therefore unlikely to "provoke a given group to hostile reaction." The jacket also could not reasonably be interpreted as "a direct personal insult" and therefore did not "violently arouse[]" any onlookers.)

[146] Cohen, 403 U.S. at 21–23 (1971) (holding that donning a shirt with the words "Fuck the Draft" in a courthouse corridor did not constitute fighting words in part because the words were not directed at anyone and people could have averted their eyes if they were offended); Purtell v. Mason, 527 F.3d 615, 625–26 (7th Cir. 2008) (holding that a family engaged in a neighborhood feud did not engage in fighting words when it displayed decorative tombstones describing its neighbors' deaths); United States v. Poocha, 259 F.3d 1077, 1082 (9th Cir. 2001) (holding that yelling "fuck you" at a police officer did not qualify as fighting words); Woods v. Eubanks, 25 F.4th 414, 425 (6th Cir. 2022) (holding that profanities like "fucking flyboy" directed at police officers were not fighting words); Cannon v. City and County of Denver, 998 F.2d 867,

Instead, the Court has steadily narrowed the doctrine's scope. In *Terminiello v. Chicago*, the Court rejected an expansive definition of "breach of the peace" that included speech merely stirring anger or unrest.[147] In *Edwards v. South Carolina*, the Court overturned convictions arising from peaceful protest, emphasizing that the expression of unpopular views does not become unprotected simply because it provokes hostility.[148] And in *Texas v. Johnson*, the Court held that flag burning, however offensive, constituted a "generalized expression of dissatisfaction" rather than a "direct personal insult or an invitation to exchange fisticuffs," as the fighting-words doctrine requires.[149] By the early 1970s, even members of the Court acknowledged the doctrine's practical dormancy. Justice Blackmun, joined by Chief Justice Burger, observed that the Court had come to pay little more than "lip service" to *Chaplinsky*.[150] Although some scholars have proposed revisions that might restore the doctrine as a meaningful First Amendment exception,[151] as a matter of positive law it remains exceedingly narrow and rarely operative.[152]

---

873–74 (10th Cir. 1993) (holding that signs reading "the killing place" carried outside an abortion clinic did not qualify as fighting words); Buffkins v. City of Omaha, 922 F.2d 465, 472 (8th Cir. 1990) (holding that calling a police officer an "asshole" did not qualify as fighting words). But see State v. Robinson, 82 P.3d 27, 31 (Mont. 2003) (holding that the words "fucking pig" and "fuck off asshole" directed at a police officer qualified as fighting words).

[147] 337 U.S. 1, 3, 5 (1949) (quoting the trial court and holding unconstitutional a jury instruction that defined a "breach of the peace" as any "misbehavior which violates the public peace and decorum" and "stirs the public to anger, invites dispute, brings about a condition of unrest, or creates a disturbance, or if it molests the inhabitants in the enjoyment of peace and quiet by arousing alarm.");

[148] 372 U.S. 229, 238 (1963) (overturning the convictions of Black students who protested against racial discrimination because peaceful expression of unpopular views does not constitute fighting words, even if such expression leads to counterprotests)

[149] 491 U.S. 397, 409 (1989) (finding that flag burning, however offensive to some, was a "generalized expression of dissatisfaction with the policies of the Federal Government," and not a "direct personal insult or an invitation to exchange fisticuffs," as fighting words doctrine requires)

[150] Gooding v. Wilson, 405 U.S. 518, 537 (1972) (Blackmun, J., dissenting).

[151] *See, e.g.*, Michael J. Mannheimer, *Fighting Words Doctrine*, 93 COLUM. L. REV. 1527, 1527–28 (1993) (discussing multiple interpretations of the constitutional effects of Chaplinsky's version of fighting words)

[152] In fact, even when addressing the fighting words cases, the Court didn't cite any part of Chaplinsky's definition but concluded that defendants' speech did not qualify as fighting words. *See* Edwards v. South Carolina, 372 U.S. 229, 236 (1963) (holding, without any analysis, that "the record is barren of any evidence of fighting words" (internal citations omitted)); Cox v. Lousiana, 379 U.S. 536, 551 (1965) (holding, without any discussion, that the record did not present "any evidence here of 'fighting words'" (citations omitted)); Norwell v. City of Cincinnati, 414 U.S. 14, 16 (1973) (holding, without analysis, that "it is clear that there was no abusive language or fighting words"); Hess v. Indiana, 414 U.S. 105, 107–08 (1973) (holding that the defendant's speech did not qualify as fighting words because it was not "directed to any person or group in particular" and, although offended, the hearer "stated he did not interpret the expression as being directed personally at him, and the evidence is clear that the appellant had his back to the sheriff at the time"); Cox Broad. Corp. v. Cohn, 420 U.S. 469, 495 (1975) (holding, without analysis, that the challenged speech "contains none of the indicia of . . . 'fighting' words, which are no essential part of any exposition of ideas, and are of such slight social value as a step to truth that any benefit that may be derived from them is clearly outweighed by the social interest in order and morality" (internal citation omitted)); Mahanoy Area Sch. Dist. v.B.L., 594 U.S. 180, 191 (2021) (holding, without analysis, that the defendant's speech "did not amount to fighting words"). In those that did, the Court only the "breach of peace" half of the Chaplinsky definition. See Street v. New York, 394 U.S. 576, 592 (1969) ("[W]e cannot say that appellant's remarks were so inherently inflammatory as to come within that small class of 'fighting words' which are likely to provoke the average person to retaliation, and thereby cause a breach of the peace . . . ."); Bachellar v. Maryland, 397 U.S. 564, 567 (1970) ("Clearly the [challenged speech] was not within that small class of 'fighting words' that, under Chaplinsky . . . are 'likely to provoke the average person to retaliation, and thereby cause a breach of the peace . . . .'" (citation omitted)). Futhermore, when applying the full Chaplinsky definition, the Court did not assess the defendants' speech but held that the challenged ordinance was either overbroad or vague when compared to the full Chaplinsky definition; *See* Gooding v.

Against this background, it is unlikely that such AI-generated speech would fall within an existing exception to First Amendment protection. Even when individualized and degrading, the fighting-words doctrine requires a narrow set of conditions that the Court has been unwilling to recognize outside paradigmatic face-to-face confrontations. The mediated and artificial character of AI-generated expression[153] further weakens any claim that such speech is likely to provoke the kind of immediate violent reaction the doctrine presupposes.

## PART III. First Amendment Protects Platform From Content Moderation

This Part argues that content-moderation regulations are likely to be held unconstitutional under the First Amendment because existing doctrine protects editorial autonomy, and mainstream platform governance scholarship has long understood moderation as an exercise of private editorial judgment by private governors. It includes three sections. The first reconstructs the First Amendment's two foundational commitments, the free circulation of information and speaker autonomy, through founding-era dictionaries. The second traces free speech jurisprudence from the 1940s to the present, showing that beginning in the 1970s the Court increasingly grounded constitutional protection in editorial autonomy, understood as control over authorship, selection, structure, expressive identity, and attribution. The third applies this doctrinal development to platform content moderation, emphasizing that once social media platforms are understood as private governors exercising editorial discretion, regulations targeting moderation practices are presumptively unconstitutional.

### A. The Two values of First Amendment

Both the text and the philosophical background of the Free Speech Clause protect two values essential to democratic self-rule: the circulation of information and the autonomy of the speaker. Textually, the Clause provides that "Congress shall make no law … abridging the freedom of speech, or of the press."[154] Although scholars have long debated the precise meaning of "speech,"[155] definitions of key terms in founding-era dictionaries point to an idea of speech organized around two values: the autonomy to decide what one will say, and the liberty to convey that expression to others.

---

Wilson, 405 U.S. 518, 524–25 (1972) (holding that the challenged statute is unconstitutionally vague and overbroad when compared to the full Chaplinsky definition).
[153] *See* Lukas Holbling, Sebastian Maier & Stefan Feuerriegel, *A Meta-Analysis of the Persuasive Power of Large Language Models*, 15 SCIENTIFIC REPORTS 43818 (2025) (summarizing that the persuasive power of large language models are contingent on and mediated by many factors).
[154] U.S. Const. amend. I.
[155] 1 RODNEY A. SMOLLA, *Smolla And Nimmer on Freedom of Speech* § 1:11 (2016) ("One can keep going round and round on the original meaning of the First Amendment, but no clear, consistent vision of what the framers meant by freedom of speech will ever emerge."); David Lat, *Justice Scalia, Originalism, Free Speech and the First Amendment*, ABOVE THE LAW (Nov. 22, 2016), ("Free speech has been kind of a desert when it comes to originalism." (quoting Michael McConnell)); *see also* Jack M. Balkin, *Nine Perspectives on* Living Originalism, 2012 U. ILL. L. REV. 815, 837 (2012) ("The abstract language of the First Amendment left unresolved differing views about the meaning of freedom of speech and press; these disputes would break out into the open later on in the 1790s . . . ."); Robert H. Bork, *Neutral Principles and Some First Amendment Problems*, 47 IND. L.J. 1, 20 (1971) ("The law has settled upon no tenable, internally consistent theory of the scope of the constitutional guarantee of free speech.")

More specifically, founding-era dictionaries catalogued multiple meanings of "speech," including: (1) an articulate utterance; (2) the expression of thoughts or ideas; (3) anything spoken; (4) talk or discourse; (5) an oration; (6) the act of speaking; (7) a declaration of thoughts; (8) a conveyance from one person's mind to another; and (9) the power of articulate utterance, that is, the power of expressing thoughts by words.[156] Three elements recur throughout these definitions: power, articulate utterance, and the expression of thought. "Power" denotes the capacity to influence, command, or control.[157] "Articulate utterance" refers to the externalization of thought through spoken or written form, conveyed distinctly and intelligibly so as to be heard or read by others.[158] "Expressing thought" means representing mental content through words, symbols, or other signs, thereby making private reflection publicly accessible.[159] Taken together, these elements presuppose an agent capable of originating expression and communicating it to others into the public sphere, with both the speaker and the channels of transmission remaining unencumbered.

The philosophical traditions merely elaborate what the text already implies. Justice Oliver Wendell Holmes, for example, articulated this understanding in his dissent in *Abrams v. United States*, where he argued that "when men have realized that time has upset many fighting faiths, they may come to believe … that the ultimate good desired is better reached by free trade in ideas—that the best test of truth is the power of the thought to get itself accepted in the competition of the market."[160] He continued:

---

[156] *See Speech*, 2 JOHN ASH, The New and Complete Dictionary of the English Language (London, Edward Dilly, Charles Dilly & R. Baldwin 1775) ("articulate utterance," "expressing thoughts," "talk," "any thing spoken"); *Speech*, 2 NATHAN BAILEY, THE NEW UNIVERSAL ETYMOLOGICAL ENGLISH DICTIONARYI (London, T. Waller, 4th ed. 1756) ("conveyance of one man's mind to another"); *Speech*, JAMES BARCLAY, A COMPLETE AND UNIVERSAL ENGLISH DICTIONARY (London, J.F. & C. Rivington et al. 1792) ("expressing our thoughts or ideas"); *Speech*, THOMAS DYCHE & WILLIAM PARDON, A NEW GENERAL ENGLISH DICTIONARY (London, Toplis, Bunney & J. Mozley, 18th ed. 1781) ("conveyance of one man's mind to another"); *Speech*, SAMUEL JOHNSON, A DICTIONARY OF THE ENGLISH LANGUAGE (London, J.F. & C. Rivington et al., 10th ed.1792) ("articulate utterance," "expressing thoughts," "oration"); *Speech*, WILLIAM PERRY, THE ROYAL STANDARD ENGLISH DICTIONARY (Worcester, 1788) ("talk," "articulate utterance"); *Speech,* 2 THOMAS SHERIDAN, A COMPLETE DICTIONARY OF THE ENGLISH LANGUAGE (London, Charles Dilly, 3d ed. 1790) ("articulate utterance," "expressing thoughts," "any thing spoken," "talk"); *Speech*, 2 NOAH WEBSTER, AN AMERICAN DICTIONARY OF THE ENGLISH LANGUAGE (New York, S. Converse 1828) ("expressing thoughts," "talk," "any declaration of thoughts").
[157] *Power*, BARCLAY, *Id.*; *Power,* SAMUEL JOHNSON, *Id.*.
[158] *Utterance*, JOHNSON, *Id*; *Utterance*, WEBSTER, *Id* ("pronunciation; manner of speaking . . . emission from the mouth; vocal expression"); *Utterance*, SHERIDAN, *Id* ("pronunciation, manner of speaking . . . vocal expression, emission from the mouth"); *Utterance*, ASH, *Id* ("[p]ronunciation, vocal expression"); *Utterance*, PERRY, *Id* ("pronunciation"); *see also Utterance*, DYCHE & PARDON, *Id* ("Speech, or the way or mode of speaking"); *Utterance*, Barclay, *Id* ("manner or power of speaking"). *Articulate*, DYCHE & PARDON, *id*; *see also Articulate*, JOHNSON, *Id* ("[d]istinct"); *Articulate*, ASH, *Id* ("distinct"); *Articulate,* SHERIDAN, *Id* ("[d]istinct"); *Articulate*, WALKER, *Id* ("[d]istinct"). *Articulate*, BARCLAY, *Id*; *see also Articulate*, WEBSTER, *Id* ("articulation of the organs of speech").
[159] *See Express*, ASH, *Id* ("to represent in words," "to shew [sic] or make known in any manner"); *Express*, BARCLAY, *Id* ("to represent in words, or by any of the imitative arts"); *Express*, DYCHE & PARDON, *Id* ("to speak or declare by word or writing"); *Express*, Johnson, *Id* ("[t]o represent by the imitative arts," "[t]o represent in words," "[t]o show or make known in any manner" "to exhibit by language"); *Express*, SHERIDAN, ("[t]o represent by any of the imitative arts," "to represent in words"); *Express*, WALKER, *Id* ("[t]o represent by any of the imitative arts," "to represent in words"); *Express*, WEBSTER, *Id* ("[t]o represent or show by imitation or the imitative arts," "[t]o show or make known").
[160] 250 U.S. 616, at 629 (1919) (Holmes, J., dissenting).

> "The Constitution is an experiment, as all life is an experiment. Every year, if not every day, we have to wager our salvation upon some prophecy based upon imperfect knowledge … [W]e should be eternally vigilant against attempts to check the expression of opinions that we loathe and believe to be fraught with death, unless they so imminently threaten immediate interference with the lawful and pressing purposes of the law that an immediate check is required to save the country."[161]

On this view, suppressing ideas is not merely regulatory overreach but a disruption of the democratic experiment itself. Since *Abrams*, courts have repeatedly invoked this logic to justify strict scrutiny for content-based restrictions[162] and to resolve disputes involving defamation liability,[163] rights of reply in newspapers,[164] denial of trademark registration for offensive marks,[165] compelled advertising,[166] student-group funding,[167] access to public forums,[168] and protest activity. The purpose is straightforward: to preserve an environment in which individuals may freely exchange ideas.

C. Edwin Baker challenges this epistemic optimism underlying free information exchange.[169] In his view, individuals do not process information in the rational and neutral manner the marketplace metaphor presumes.[170] Instead, they selectively perceive, selectively expose themselves to, and selectively retain information, while ignoring or discrediting material inconsistent with their existing beliefs.[171] On this account, any confidence that the marketplace reliably delivers the "best" truths should, in Baker's words, be "eviscerated."[172] Nor is the communicative environment itself even-handed. Mass media structures and unequal access entrench dominant perspectives, rendering reliance on circulation alone "improperly biased in favor of presently dominant groups."[173] In this environment, even fallibilist defenses falter, as the mere acknowledgment of human error supplies neither an account of which purposes are to be served nor a fair standard for judging outcomes.

---

[161] *Id.*
[162] Reed v. Town of Gilbert, 576 U.S. 155, 171 (2015) (Breyer, J., concurring).
[163] New York Times Co. v. Sullivan, 376 U.S. 254 (1964).
[164] Miami Herald Publ'g Co. v. Tornillo, 418 U.S. 241, 248 (1974) (rejecting notions of a right of access to newspapers, while noting: "The contentions of access proponents will be set out in some detail. It is urged that at the time the First Amendment to the Constitution was ratified in 1791 as part of our Bill of Rights the press was broadly representative of the people it was serving. While many of the newspapers were intensely partisan and narrow in their views, the press collectively presented a broad range of opinions to readers. Entry into publishing was inexpensive; pamphlets and books provided meaningful alternatives to the organized press for the expression of unpopular ideas and often treated events and expressed views not covered by conventional newspapers. A true marketplace of ideas existed in which there was relatively easy access to the channels of communication.").
[165] Matal v. Tam, 582 U.S. 218, 243 (2017) ("Refusing to confer an even greater benefit, we held, did not upset the marketplace of ideas and did not abridge the union's free speech rights"); *Id.* at 1766 (Kennedy, J., concurring) ("The First Amendment's viewpoint neutrality principle protects more than the right to identify with a particular side. It protects the right to create and present arguments for particular positions in particular ways, as the speaker chooses. By mandating positivity, the law here might silence dissent and distort the marketplace of ideas … J Holmes' reference to the 'free trade in ideas' and the 'power of … thought to get itself accepted in the competition of the market . . .,' was a metaphor. In the realm of trademarks, the metaphorical marketplace of ideas becomes a tangible, powerful reality. Here that real marketplace exists as a matter of state law and our common-law tradition, quite without regard to the Federal Government.").
[166] *See* United States v. United Foods, Inc., 533 U.S. 405, 409 (2001) (In deciding whether mushroom growers could be forced to support generic mushroom advertising, the Court observed, "Thus, government statements (and government actions and programs that take the form of speech) do not normally trigger the First Amendment rules designed to protect the marketplace of ideas.").
[167] Rosenberger v. Rector & Visitors of Univ. of Virginia, 515 U.S. 819, 847 (1995) (O'Connor, J., concurring) ("It is clear that the University has established a generally applicable program to encourage the free exchange of ideas by its students, an expressive marketplace that includes some 15 student publications with predictably diver- gent viewpoints.").
[168] For the various public places, such as utility pole posting, *See* Members of City Council of City of Los Angeles v. Taxpayers for Vincent, 466 U.S. 789, 804 (1984) ("[T]here are some purported interests — such as a desire to suppress support for a minority party or an unpopular cause, or to exclude the expression of certain points of view from the marketplace of ideas — that are so plainly illegitimate that they would immediately invalidate the rule."), for abortion protests: McCullen v. Coakley, 573 U.S. 464, 476 (2014) (failing to "'preserve an uninhibited marketplace of ideas in which truth will ultimately prevail'") (quoting FCC v. League of Women Voters of Cal,, 468 U.S. 364, 377 1(984)).
[169] *See* C. EDWIN BAKER, HUMAN LIBERTY AND FREEDOM OF SPEECH (Oxford Univ. Press 1989).

Yet Baker's critique ultimately clarifies rather than undermines the Free Speech Clause by illuminating its second core commitment: if public discourse cannot be expected to end in correct answers, then the value of speech must rest in the autonomy of the speaker rather than in the epistemic promise of the marketplace.[174] Expression warrants protection not because it guarantees desirable outcomes, but because it honors the speaker's authority to decide what to say, how to say it, and whether to speak at all. Speech thus becomes an end in itself. By securing that authority, the First Amendment enables self-fulfillment, understood as the formation and realization of identity through expression, and guarantees participation in change, ensuring that every member of society, whether disgruntled or visionary, retains the capacity to help shape the community's future. What eventually matters is not that every valuable idea be spoken or that society converges on correct answers, but that no one be denied the right to speak at all.

Alexander Meiklejohn offers an additional supporting account through his democratic-process theory. There, he argues that a people cannot govern themselves unless they are fully informed, and they cannot be fully informed unless all perspectives on matters of public concern are heard.[175] Citizens must therefore have equal access to the information necessary for collective decision-making.[176] That information, moreover, extends beyond immediate political debate to include literature, philosophy, religion, and the arts, since restricting such expression would stunt the faculties of reason and imagination on which self-government depends.[177] In sum, the Clause reflects two values: the free circulation of ideas and the autonomy of individuals in expressing their thoughts.

---

[170] *Id* at 12-17.
[171] *Id*.
[172] *Id* at 15.
[173] *Id*.
[174] *Id* at 20.
[175] *See generally* ALEXANDER MEIKLEJOHN, FREE SPEECH AND ITS RELATION TO SELF-GOVERNMENT (Lawbook Exch. 2000)
[176] *Id* at 26. *See also* Alexander Tsesis, *Free Speech Constitutionalism*, 2015 U. ILL. L. REV. 1015, 1035-36 (2015) (describing how the theory distinguishes between "political and private speech").
[177] For early critique, *see supra* note 31 Emerson (praising Meiklejohn's focus on self-government but points out the unmanageable sweep of including the arts and sciences as "political."); *see also* Robert Post, *Participatory Democracy and Free Speech*, 97 VA L. REV 477 (2011) (stressing that democratic legitimation requires protecting speech that builds the conditions for deliberation, but criticizing Meiklejohn's tendency to treat all cultural development as instrumental to politics).

## B. Editorial Autonomy Trumping Over Information Circulation

Since the 1970s, the Court has shifted from protecting the free circulation of information to a more robust commitment to preserving editorial judgment, expressive identity, and control over authorship. This section lays the doctrinal foundation for Section C's central implication: once platform moderation is understood as an exercise of editorial choice, regulations targeting those choices are likely to be found unconstitutional under the First Amendment.

### 1. Protecting Information Circulation

From the 1940s through the 1970s, the Court sought to preserve free circulation of information by keeping public forums open, insulating distribution channels from formal and informal state pressures that induce self-censorship, and, in limited contexts, preventing the monopolization of scarce communicative resources.

### (1) Keeping Forums Open

The Court initially sought to keep the communicative forums open. In *Marsh v. State of Alabama,* a Jehovah's Witness stood on the sidewalk of the company-owned business block in Chickasaw, a company town outside Mobile, near the post office, and distributed religious pamphlets.[178] Chickasaw was owned by Gulf Shipbuilding but operated, in every practical respect, like any other town.[179] Its streets, sidewalks, stores, and post office were open to and regularly used by the public, with company roads connecting directly to a four-lane highway.[180] The company had posted a notice barring solicitation without written permission and informed the Witness that no permit would be issued.[181] She refused to leave, insisting that the rule could not constitutionally bar her leafleting.[182] A company-paid deputy sheriff arrested her, and the state charged her under Alabama's trespass statute for remaining on the premises after warning.[183]

The Court ruled in her favor.[184] It held that a State may not invoke trespass law to criminally punish a person for distributing religious literature on the sidewalks of a company town that is open and operated like any ordinary municipality.[185] The fact that Gulf Shipbuilding held title to the town did not confer absolute power over its operation.[186] To the contrary, "the more [an owner], for his advantage, opens up his property for use by the public in general, the more do his rights become circumscribed by the statutory and constitutional rights of those who use it."[187] The Court analogized company towns to other privately owned facilities that perform public functions: "owners of privately held bridges, ferries, turnpikes and railroads may not operate them as freely as a farmer does his farm."[188] The constitutional inquiry turned not on formal ownership, but on functional role. Many people, the Court recognized, live in company towns, and like all citizens, they must make decisions affecting the welfare of their communities and the nation.[189] To do so responsibly, "their information must be uncensored."[190] Therefore, when private property becomes the site of collective life, it must yield to the free distribution of information essential to civic participation and self-government.

---

[178] 326 U.S. 501, 503 (1946).
[179] *Id.*
[180] *Id.*
[181] *Id.* at 504.
[182] *Id.*
[183] *Id.*
[184] *Id* at 506-509.
[185] *Id.*
[186] *Id* at 506.
[187] *Id.*

The logic of *Marsh* was not confined to company towns. In *Food Employees v. Logan Valley Plaza, Inc.*, the Court considered whether Pennsylvania could use its trespass law to enjoin peaceful union picketing in the privately owned Logan Valley Mall.[191] The issue, as the Court framed it, was whether peaceful picketing of a business enterprise located within a shopping center may be barred as an unconsented invasion of the owner's property rights.[192]

Here, the Court again ruled in favor of the speakers.[193] It held that, for First Amendment purposes, a shopping center that functions as the community's business block and is freely accessible to the public must be treated "in substantially the same manner."[194] The Court began from the presumption "that peaceful picketing carried on in a location open generally to the public is, absent other factors involving the purpose or manner of the picketing, protected by the First Amendment."[195] That picketing involves both speech and conduct, placards and patrolling, did not strip it of protection.[196] As the Court emphasized, "no case decided by this Court can be found to support the proposition that the nonspeech aspects of peaceful picketing are so great as to render the provisions of the First Amendment inapplicable to it altogether."[197]

What proved decisive was the location of the protest. Expression, the Court reasoned, is meaningful only if speakers can reach their intended audience.[198] Forcing the picketers to the berms outside the mall would have rendered their message practically invisible.[199] As the Court explained, the placards could be read only "at a distance so great as to render them virtually indecipherable, or while the prospective reader is moving by car from the roads onto the mall parking areas."[200] If customers could not read the placards, and if distributing handbills to motorists on busy highways was both impractical and hazardous, then the right of expression was nullified in effect.[201] Protecting expression therefore required not only recognizing picketing as speech, but also securing a place for it where the message could be practically conveyed to its intended audience.[202]

---

[188] *Id.*
[189] *Id.* at 507.
[190] *Id.* at 508.
[191] 391 U.S. 308 (1968).
[192] *Id* at 309.
[193] *Id.*
[194] *Id* at 324.
[195] *Id* at 313.
[196] *Id* at 314.
[197] *Id.*
[198] *Id* at 322.
[199] *Id.*
[200] *Id.*
[201] *Id.*

That expansive logic, however, did not endure. Only four years later, in *Lloyd Corp. v. Tanner*,[203] the Court narrowed the obligations imposed on private property owners. The case involved anti–Vietnam War protesters distributing leaflets inside a privately owned shopping mall.[204] Unlike in Logan Valley, their protest bore no relation to the businesses within the mall.[205] They were not addressing labor conditions, wages, or even the mall's tenants, but the policies of the federal government.[206] The Court seized on this distinction and held that the First Amendment did not require the mall's owners to open their property as a public forum.[207] Where "adequate alternative channels" of communication exist, the Court concluded, the right to expression does not entail the right to use another's property.[208] Subsequent cases have consistently confined *Marsh* to its narrow facts, treating it as an exception that applies only when private property is the functional equivalent of an entire municipality.[209]

This commitment to openness is not confined to physical locations alone. In *Packingham v. North Carolina*,[210] the Court extended the principle of open access from traditional civic spaces to digital ones. The question was whether a State could categorically bar an entire class of people from accessing social media platforms.[211] The Court answered in the negative, reaffirming that "a fundamental principle of the First Amendment is that all persons have access to places where they can speak and listen, and then, after reflection, speak and listen once more."[212] "Cyberspace," and "social media in particular," Justice Kennedy explained, now constitute the "vast democratic forums of the Internet."[213] They are the modern sites for "knowing current events," seeking employment, petitioning representatives, and engaging in public life,[214] offering a "relatively unlimited, low cost capacity for communication of all kinds."[215] To foreclose access to these platforms is therefore to bar entry to what functions as the contemporary public square and, in doing so, to "prevent the user from engaging in the legitimate exercise of First Amendment rights."²¹⁶ Of course, this commitment to openness does not mean that all forums must remain equally accessible to all persons in all circumstances. Subsequent cases have largely declined to extend *Packingham* to invalidate restrictions imposed on particular classes of offenders, especially individuals convicted of sexual offenses, when access limitations are imposed as individualized conditions of probation, parole, or supervised release rather than as categorical statutory bans.²¹⁷

---

[202] *Id.*
[203] 407 U.S. 551 (1972)
[204] *Id.*
[205] *Id* at 552.
[206] *Id.*
[207] *Id.* at 573 ("we must decide whether ownership of the Center gives petitioner unfettered discretion to determine whether or not it will be used as a public forum…in our view, the circumstance that the property rights to the premises where the deprivation of liberty, here involved, took place, were held by others than the public, is not sufficient.").
[208] *Id* at 567.
[209] *See* e.g., Hudgens v. NLRB, 424 U.S. 507, 518 (1976) (distinguishing *Marsh* as limited to its facts); Jackson v. Metropolitan Edison Co., 419 U.S. 345, 352–53 (1974) (private utility not equivalent to municipality); Flagg Bros., Inc. v. Brooks, 436 U.S. 149, 159–60 (1978) (*Marsh* an "extraordinary" case); Manhattan Cmty. Access Corp. v. Halleck, 587 U.S. ___, slip op. at 10 (2019) (*Marsh* is a "narrow exception"); Brentwood Acad. v. Tenn. Secondary Sch. Athletic Ass'n, 531 U.S. 288, 296 (2001) (placing *Marsh* within a narrow public-function category).
[210] 582 U.S. 98 (2017).
[211] *Id* at 101.
[212] *Id* at 104.
[213] *Id* at 104.
[214] *Id* at 107.
[215] *Id* at 104.

**(2) Make Free Information Circulation**

The Court has also sought to preserve the free circulation of information, primarily along two dimensions: first, by protecting the free distribution of expressive materials; and second, by preventing the monopolization of scarce communicative resources. Both dimensions, though never overturned, were consistently refined and limited to the specific historical and technological contexts from which the cases arose.

For the first dimension, the protection of free distribution, *Smith v. California*²¹⁸ and *Bantam Books v. Sullivan*²¹⁹ offer two examples. In *Smith*, the appellant was a Los Angeles bookseller convicted under a city ordinance that made it a crime "for any person to have in his possession any obscene or indecent writing [or] book in any place of business where books are sold or kept for sale."²²⁰ The ordinance imposed strict liability.²²¹ As a result, Smith was sentenced to jail after an obscene book was found in his shop, without any proof that he knew its contents.²²² Smith objected that this elimination of scienter conflicted with the Constitution.²²³ The Court agreed. It reasoned that while obscenity itself lies outside the protection of the First Amendment,²²⁴ the State may not enforce that limit through a rule that strangles the circulation of protected works.²²⁵ By eliminating scienter, California had imposed "a severe limitation on the public's access to constitutionally protected matter," forcing booksellers into self-censorship.²²⁶ A bookseller facing absolute criminal liability would predictably narrow his inventory to the few titles he could personally inspect or avoid controversial works altogether, a demand the Court described as "altogether unreasonable … and near to omniscience."²²⁷ That burden would inevitably travel to the public, "for by restricting him the public's access to reading matter would be restricted."²²⁸ In effect, compelled self-censorship of booksellers would dry up the channels of circulation. It is not merely a single obscene volume at stake, but rather the integrity of the very channel through which the public remains informed.

---

²¹⁶ *Id* at 108.
²¹⁷ *See* United States v. Finnell, No. 22-13892, 2023 WL 6577444 1 (11th Cir. Oct. 10, 2023) (declined to extend *Packingham* because lifetime supervised-released conditions restricting internet use were part of sentencing for child-pornography offenses.)
²¹⁸ 361 U.S. 147 (1959).
²¹⁹ 372 U.S. 58 (1963).
²²⁰ *Smith* 361 U.S. at 148.
²²¹ *Id.*
²²² *Id.*
²²³ *Id.*
²²⁴ *Id* at 152 ("We have held that obscene speech and writings are not protected by the constitutional guarantees of freedom of speech and the press.")
²²⁵ *Id* at 155.
²²⁶ *Id* at 153.
²²⁷ *Id.*

*Bantam Books* extended this protection of circulation from formal statutory prohibitions to informal, de facto systems of suppression.²²⁹ In *Bantam Books*, four New York paperback publishers, including Bantam and Dell, distributed their titles in Rhode Island through Max Silverstein & Sons, the State's exclusive wholesaler.²³⁰ Rhode Island had created a Commission "to encourage morality in youth," whose practice was to send official notices on Commission letterhead to distributors declaring certain books and magazines "objectionable" for sale or display to minors.²³¹ The notices thanked distributors in advance for their "cooperation," reminded them of the Commission's duty to recommend obscenity prosecutions, and copied local police departments.²³² Silverstein received approximately thirty-five such notices.²³³ In response, he refused to fill orders for the listed books, directed field representatives to remove unsold copies from retailers, and returned them to publishers.²³⁴ As he explained, he complied "rather than face the possibility of some sort of a court action" by a "duly authorized organization."²³⁵

The Court again ruled for the speakers.²³⁶ It emphasized that it was immaterial that the Commission lacked formal power to ban books, for "the record amply demonstrates that [it] deliberately set about to achieve the suppression of publications deemed 'objectionable' and succeeded in its aim."²³⁷ The notices were "phrased virtually as orders," followed by police visits, and "stopped the circulation of the listed publications ex proprio vigore."²³⁸ Moreover, "people do not lightly disregard public officers' thinly veiled threats to institute criminal proceedings against them."²³⁹ What the State styled as mere advice thus operated, in substance, as a form of effective state regulation eliminating the safeguards of the criminal process.²⁴⁰ And because the Commission's system provided no notice, no hearing, and no judicial review, but instead functioned under a vague mandate that left distributors guessing whether a book was obscene or merely "harmful to juvenile morality,"²⁴¹ the resulting uncertainty, backed by the threat of prosecution, suppressed circulation altogether and was therefore unconstitutional.²⁴² Many subsequent cases refine this logic, clarifying that state action, whether by seizure,[243] vague procedure,[244] economic burden,[245] or indirect barrier,[246] violates the First Amendment when it chills the circulation of information.[247]

---

[228] *Id* at 154.
[229] 372 U.S. at 59.
[230] *Id* at 61.
[231] *Id* at 59.
[232] *Id* at 62.
[233] *Id* at 61.
[234] *Id* at 63.
[235] *Id*.
[236] *Id* at 72.
[237] *Id* at 67.
[238] *Id* at 68.
[239] *Id* at 68.
[240] *Id* at 69-70.
[241] *Id* at 71.
[242] *Id*.

For the second dimension, the prevention of monopolization of scarce communicative resources, the Court aimed to secure public access to a diverse range of views, including those that compete or conflict with one another. The governing principle is that, as the Court has repeatedly held, "it is the right of the viewers and listeners, not the right of the broadcasters, which is paramount" in safeguarding access to public debate.[248] *Red Lion Broadcasting Co. v. FCC*[249] demonstrated this. In *Red Lion*, Red Lion Broadcasting aired a fifteen-minute program attacking Fred J. Cook's book Goldwater—Extremist on the Right.[250] When Cook requested reply time pursuant to the Federal Communications Commission's fairness doctrine, the station refused.[251] The Court upheld the FCC's order, reasoning that the doctrine enhanced, rather than abridged, First Amendment freedoms.[252] The broadcast spectrum, the Court emphasized, is a scarce public resource whose allocation cannot be left entirely to private control, for private allocation had historically produced not robust debate but chaos, "a cacophony of competing voices, none of which could be clearly and predictably heard."[253] A broadcaster, as the holder of public resources "of considerable and growing importance,"[254] was therefore properly treated as a fiduciary for the community, obligated to serve the public "convenience, interest, or necessity," the fulfillment of which required presenting views representative of the community, not merely those the broadcaster itself chose to endorse.[255]

---

[243] *See* Marcus v. Search Warrant, 367 U.S. 717 (1961) (mass seizures without prior adversary hearings risked sweeping up protected works and "impose a prior restraint on distribution." When whole channels of circulation are blocked by seizure, the effect is to silence protected speech, not just obscene material)

[244] *See* A Quantity of Books v. Kansas, 378 U.S. 205 (1964) (blocking entire inventories suppresses circulation of non-obscene books as well; reinforcing *Smith* concern that overbroad regulation collapses distribution itself); Freedman v. Maryland, 380 U.S. 51 (1965) (about prior restraint; procedural safeguards are necessary to prevent suppression of protected films. Echoing *Bantam Books* that administrative schemes lacking notice and judicial review would chill circulation); Interstate Circuit Inc. v. City of Dallas, 390 U.S. 676 (1968) (court holding that vague standards like "morally unsuitable" chill distributors and exhibitors, forcing them to guess what's forbidden and leading to over-censorship)

[245] *See* Grosjean v. American Press Co., 297 U.S. 233 (1936) (Court striking down Louisiana's special ta on high-circulation newspapers, seeing it as a penalty on press distribution); Minneapolis Star & Tribune Co. v. Minn. Comm'r of Revenue, 460 U.S. 575 (1983) (Court invalidating a tax on paper and ink used by large newspapers; even facially neutral taxes that single out press distribution was unconstitutional); Arkansas Writers' Project v. Ragland, 481 U.S. 221 (1987) (striking down sales tax on some magazines but not others; Court ruled that such discriminatory economic burdens distort circulation).

[246] *See* Lamont v. Postmaster General, 381 U.S. 301 (1965).

[247] *See e.g.,* Watchtower Bible and Tract Society of New York, Inc. v. Village of Stratton, 536 U.S. 150 (2002) (Court striking down a permit requirement for door to door advocacy because it burdened the "places and channels" where civic conversation naturally occur); City of Ladue v. Gilleo, 512 U.S. 43 (1994) (The City banned residential signs to promote aesthetics; the Court said that wiped out an entire "channel of communication" uniquely suited for political self-expression).

[248] *See* FCC v. Sanders Bros. Radio Station, 309 U.S. 470, 475 (1940); FCC v. Allentown Broadcasting Corp. 349 U.S. 358, 361-362 (1955).

[249] 395 U.S. 367 (1969).

[250] *Id* at 371.

[251] *Id* at 372.

[252] *Id* at 370-371.

[253] *Id* at 376.

[254] *Id* at 399.

[255] *Id.*

Formally, *Red Lion* remains good law. In practice, however, the Court has largely confined it to its historical and technological setting, declining to extend its scarcity-based reasoning beyond traditional broadcast media. In *Amann v. Clear Channel Communications*,[256] for example, courts refused to impose Red Lion–style fiduciary duties on radio broadcasters in tort contexts.[257]

This retreat is most explicit in *Turner Broadcasting System, Inc. v. FCC*,[258] where the Court grounded First Amendment analysis in the realities of technological architecture. In *Turner*, the Court emphasized that the relaxed standard of scrutiny applied in broadcast cases rests on the unique physical limitations of the electromagnetic spectrum, not on market power, concentration, or alleged dysfunction in a speech market.[259] In the broadcast context, there were substantially more would-be speakers than available frequencies, and because simultaneous transmission over the same frequency causes signal interference, regulation was necessary simply to render broadcasting possible.[260] Those premises, however, do not transfer to cable television.[261] Cable does not suffer from the inherent physical limitations that characterize broadcasting.[262] Advances in fiber optics and digital compression technology eliminate any practical ceiling on the number of speakers, and there is no danger of physical interference when multiple speakers share the same channel.[263] For that reason, the Court deemed *Red Lion*'s scarcity-based framework "inapt," and with it the constitutional justification for treating private speakers as public fiduciaries.[264]

**2. Shifting to Editorial Autonomy**

Since the 1970s, the Court has grounded First Amendment protection in editorial autonomy, specifically in speakers' control over what to publish, which subjects to address, and how expressive materials are structured. It rejects both direct and indirect governmental efforts to compel inclusion, alter subject matter, or displace editorial discretion. This protection extends beyond discrete editorial decisions to expressive identity itself, holding that associations and speakers may define their own messages and membership without state interference or speaker-based exclusion from public debate. At the same time, the Court has guarded against state-compelled speech and false attribution, striking down laws that force speakers to carry, create, or appear to endorse messages they do not choose to express.

**(1) Protecting editorial choices for materials**

---

[256] 846 N.E.2d 95 (2006).
[257] *Id.*
[258] 512 U.S. 622 (1994).
[259] *Id* at 632.
[260] *Id* at 637.
[261] *Id.*
[262] *Id.*
[263] *Id* at 639.
[264] *Id.*

Per *Miami Herald Publishing Co. v. Tornillo*, the First Amendment protects editorial judgment over what to publish, how to publish it, and in what form.[265] Governmental compulsion that dictates inclusion, exclusion, or editorial arrangement impermissibly intrudes on constitutionally protected speech.[266] Specifically, in *Tornillo*, the appellant, a Florida newspaper had published editorials criticizing a legislative candidate.[267] Invoking the state's "right of reply" statute, the candidate demanded that the paper print his response verbatim and without charge.[268] The Herald refused, arguing that the statute "purports to regulate the content of a newspaper,"[269] leaving editors unable to know in advance which criticisms would trigger mandatory publication.

The Court agreed.[270] It reasoned that a law that forces editors to print material "which 'reason' tells them should not be published" is indistinguishable from a law forbidding them to print what they wish to publish.[271] As the Court explained, "governmental compulsion" of this kind interferes with "the crucial process" by which editors decide "the choice of material to go into a newspaper," the "size and content of the paper," and the "treatment of public issues and public officials," and is thereby unconstitutional.[272] It matters not that the statute imposed no financial burden and required no sacrifice of space. Even if compliance "would face no additional costs," the statute failed when it "intruded into the function of editors."[273]

Under the First Amendment, editors are also free to choose the subjects of the materials they publish. In *Consolidated Edison Co. v. Public Service Commission*,[274] the New York Public Service Commission had barred Consolidated Edison from enclosing in its monthly bills any discussion of "controversial issues of public policy."[275] The court held such an order unconstitutional as it altered the company's choice of subject.[276] As Justice Powell explained, "The First Amendment means that government has no power to restrict expression because of its message, its ideas, its subject matter, or its content."[277] A regulation that prohibits "public discussion of an entire topic" is a direct content based restraint.[278] If the marketplace of ideas is to remain free and open, "governments must not be allowed to choose which issues are worth discussing or debating," for allowing such control would grant the state power over "the search for political truth."[279] By "limiting the means by which Consolidated Edison may participate in the public debate on this question and other controversial issues of national interest and importance," the Commission had "struck at the heart of the freedom to speak."[280]

---

[265] 418 U.S. 241 (1974).
[266] *Id.*
[267] *Id* at 243 - 244.
[268] *Id* at 244.
[269] *Id.*
[270] *Id* at 258.
[271] *Id* at 256.
[272] *Id.* at 258.
[273] *Id.*
[274] 447 U.S. 530 (1980).
[275] *Id* at 532.
[276] *Id.*
[277] *Id* at 537.
[278] *Id.*
[279] *Id* at 538.

A speaker shall remain free to edit its message. In *Columbia Broadcasting System v. Democratic National Committee*,[281] Chief Justice Burger, joined by Justices Stewart and Rehnquist, rejected the DNC's attempt to require CBS to sell airtime for editorial advertisements.[282] They emphasized that the First Amendment does not permit the government to displace a broadcaster's editorial judgment with the demands of would-be speakers.[283] The FCC's role, as the Court explained, is to promote a diversity of sources, not to "edit" or "screen" what licensees may broadcast.[284] "Editing," so say the judges, "is what editors are for; and editing is selection and choice of material."[285] Decisions about whether to include or exclude a message, and how to allocate the finite commodity of airtime, fall squarely within the broadcaster's protected autonomy. The government may not redistribute that autonomy by granting outsiders a right to insert their own speech into someone else's program.[286] Subsequent cases confirm this principle. In *Writers Guild of America, West, Inc. v. FCC*,[287] the court held unconstitutional the FCC's role in promoting the "family viewing policy," where content was altered not through independent editorial judgment but in response to networks' anticipation of regulatory displeasure.[288] The policy pressured broadcasters to modify characters, abandon themes, and excise language, thereby distorting editorial choice through indirect compulsion.[289] Likewise, in *Barnstone v. University of Houston, KUHT-TV*, the court held that programming decisions must remain with professional editors, warning that compelled access would reduce them from decisionmakers to mere conduits for others' speech.[290]

### (2) Preserving speakers' expressive identity

Speakers, under the First Amendment, retain their expressive identity; they may freely choose to include or exclude members whose traits align with the message the group seeks to convey. State actions that overwrite, dilute, or redirect that identity violates the Free Speech clause. In *Boy Scouts of America v. Dale*,[291] BSA revoked the adult membership of James Dale after learning he was an openly gay scoutmaster and activist, asserting that his public identity contradicted the values the organization sought to instill.[292] Dale sued under New Jersey's public accommodations law.[293] The Court, in holding for BSA, stated that, "forced inclusion of an unwanted person in a group infringes the group's freedom of expressive association if the presence of that person affects in a significant way the group's ability to advocate public or private viewpoints."[294] Expressive identity, the Court emphasized, is something the association itself defines.[295] "We must give deference to an association's view of what would impair its expression," and the state may not substitute its own account of fairness.[296] As the Court put it, "The state interests embodied in New Jersey's public accommodations law do not justify such a serious intrusion on the Boy Scouts' rights to freedom of expressive association."[297] Membership, on this view, is part of the group's message. If the Boy Scouts asserts that homosexual conduct is inconsistent with the values it seeks to instill, then including an assistant scoutmaster who is an avowed homosexual would force the organization to send a message with which it does not agree.[298] Thus, "the forced inclusion of Dale would significantly affect the Boy Scouts' ability to advocate its viewpoints."[299] New Jersey's mandate would have altered the organization's own message, a power the First Amendment withholds from the state.[300]

---

[280] *Id* at 535.
[281] 412 U.S. 94 (1973).
[282] *Id* at 132.
[283] *Id* at 124-126.
[284] *Id* at 117-118.
[285] *Id* at 124.
[286] *Id.*
[287] 423 F.Supp. 1064 (1976).
[288] *Id* at 1153.
[289] *Id.*
[290] 660 F.2d 137 (1981).
[291] 530 U.S. 640 (2000).
[292] *Id* at 645.
[293] *Id.*

Any group, be it a corporation, association, union, or individual, may claim this expressive identity when speaking in public debate, without the legislature conditioning that eligibility on the government's preferred account of who the speaker is or how it should address a controversial issue. In *First National Bank of Boston v. Bellotti*,[301] Massachusetts barred corporations from spending money to influence ballot initiatives unless the issue "materially affected" their business.[302] The Court held the statute unconstitutional.[303] The First Amendment, it reasoned, protects speech rather than speakers, and the value of speech does not depend on the identity of the entity that utters it.[304] As the Court put it, "The inherent worth of the speech in terms of its capacity for informing the public does not depend upon the identity of its source."[305] The state therefore may not decide which speakers are qualified to participate in public debate, nor may it condition eligibility on corporate form or degree of stake.[306] A restriction that excludes a class of speakers from participation in a public referendum impermissibly distorts the marketplace of ideas. To keep that marketplace open, the government may not "restrict the speech of some elements of society in order to enhance the relative voice of others."[307]

Protection for expressive identity attaches when a group is engaged in producing an expressive product, articulating a message, or making choices that constitute its public identity. When an organization performs functions that are non-expressive with respect to that identity, the First

---

[294] *Id* at 648.
[295] *Id* at 650.
[296] *Id.*
[297] *Id* at 659.
[298] *Id* at 650-652.
[299] *Id* at 650.
[300] *Id* at 654.
[301] 435 U.S. 765 (1978).
[302] *Id* at 768.
[303] *Id* at 767.
[304] *Id* at 777.
[305] *Id.*
[306] *Id* at 785.
[307] *Id* at 791.

Amendment does not apply. In *Rumsfeld v. FAIR*,[308] the law schools argued that hosting military recruiters compelled them to disseminate the government's message.[309] The Court rejected that analogy outright. As Chief Justice Roberts explained, the schools "are not speaking when they host interviews and recruiting receptions."[310] Decisions about interview rooms and scheduling "lack the expressive quality of a parade."[311] Accommodating the military's message therefore neither alters nor interferes with any message the schools themselves convey; recruiting is simply a service "to assist their students in obtaining jobs," not an act of institutional self-expression.[312]

Additionally, where an association's expressive claims are diffuse, unselective, and weakly articulated, First Amendment protection does not attach. In *Roberts v. United States Jaycees*,[313] the Jaycees were a national nonprofit membership corporation, organized to promote the civic and professional development of young men.[314] They refused to admit women as full members and argued that compelled inclusion under Minnesota's public accommodations law would dilute their message.[315] The Court rejected that claim. Writing for the Court, Justice Brennan emphasized that the Jaycees were a large national organization whose local chapters were "neither small nor selective," and whose activities regularly involved "the participation of strangers."[316] Lacking the "distinctive characteristics"[317] of an expressive association, the Jaycees' membership decisions did not merit constitutional protection. Its associational choices were not integral to the deliberate cultivation and transmission of a shared moral vision in the way they were for the Boy Scouts in *Dale*, whose expressive identity was concretely articulated through the Scout Oath, the Scout Law, and the organization's hierarchical leadership structure.[318]

### (3) Guarding Against State-Attributed Speech

States may not treat an entity as the author of views it has neither adopted nor chosen to express. In *Hurley v. Irish-American Gay, Lesbian, and Bisexual Group of Boston*,[319] the Court addressed whether the South Boston Allied War Veterans Council, organizers of the city's St. Patrick's Day–Evacuation Day Parade, could be compelled under Massachusetts's public accommodations law to include among the marchers a group imparting a message the organizers did not wish to convey.[320] The Court held that it could not.[321] For, as Justice Souter explained, a parade is "a form of expression," a "public drama of social relations" that communicates ideas through words, symbols, and collective action.[322] The Council is, in itself, an expressive author: "Rather like a composer," it "selects the expressive units of the parade from potential participants."[323] Having chosen to exclude a message it did not wish to convey, the Council was entitled to "shape its expression by speaking on one subject while remaining silent on another."[324] MA's compelling it to include a member it otherwise would have altered the meaning of the parade itself by falsely attributing GLIB's message to the organizers. As the Court emphasized, GLIB's participation would "likely be perceived as having resulted from the Council's customary determination that its message was worthy of presentation and quite possibly of support as well."[325] In this sense, the Council could not be compelled to convey or endorse a message it had chosen not to express.

---

[308] 547 U.S. 47 (2006).
[309] *Id* at 53.
[310] *Id* at 64.
[311] *Id* at 49.
[312] *Id* at 64.
[313] 468 U.S. 609 (1984).
[314] *Id.* at 612.
[315] *Id* at 617.
[316] *Id* at 621.
[317] *Id* at 620.
[318] *Id* at 621-622.
[319] 515 U.S. 557 (1995)
[320] *Id.* at 558.
[321] *Id.*

Nor may the state compel an entity to carry a message with which it disagrees. In *Pacific Gas & Electric Co.*, the Court considered whether the California Public Utilities Commission could require a privately owned utility to include, in its billing envelopes, speech authored by a third party and opposed by the utility.[326] A plurality led by Justice Powell rejected the order.[327] Joined by Chief Justice Rehnquist and Justice O'Connor, the plurality explained that compelled access of this kind both "penalizes the expression of particular points of view and forces speakers to alter their speech to conform with an agenda they do not set."[328] By requiring Pacific to distribute third-party advocacy in envelopes bearing its name and address, the Commission "require[d] appellant to use its property as a vehicle for spreading a message with which it disagrees,"[329] thereby infringing the utility's "right of refraining from speaking at all."[330] Chief Justice Burger concurred, emphasizing that the case turned on "the infringement of Pacific's right to be free from forced association with views with which it disagrees."[331] States may also not compel speakers to produce expression they would otherwise refuse. In *303 Creative v. Elenis*,[332] Colorado sought to deploy its public-accommodations law to require businesses to provide goods and services to all customers regardless of race, creed, disability, sexual orientation, or other protected traits, on pain of civil penalties.[333] As applied, the law would have required the petitioner, Ms. Smith, to design wedding websites celebrating same-sex marriages in violation of her religious convictions.[334] The Court rejected that application.[335] It held that the First Amendment forbids the State from compelling such expression where the requested services are "original," "customized," and "expressive in nature," and where viewers would reasonably attribute the resulting message to the speaker herself.[336] By mandating the creation of speech the designer did not wish to provide, Colorado crossed the constitutional line. Compelled expression of this kind, the Court concluded, is an impermissible abridgment of the Free Speech Clause.

---

[322] *Id* at 568.
[323] *Id* at 574.
[324] *Id.*
[325] *Id* at 575.
[326] 475 U.S. at 4.
[327] *Id.*
[328] *Id* at 9.
[329] *Id* at 17.
[330] *Id* at 32.
[331] *Id* at 21.
[332] 600 U.S. 570 (2023).
[333] *Id* at 581.
[334] *Id* at 582.
[335] *Id* at 603.

But where the speech at issue is not reasonably attributable to the owner, compelled access or accommodation does not amount to compelled speech. In *PruneYard Shopping Center v. Robins*,[337] the Court confronted a large, privately owned shopping center open to the general public, containing dozens of stores, restaurants, walkways, and plazas.[338] A group of high school students set up a small table in a common courtyard to distribute pamphlets and collect signatures opposing a United Nations resolution.[339] Relying on a rule barring expressive activity unrelated to commerce, the shopping center ordered them to leave.[340]

The Court rejected the claim that compelled access constituted compelled speech. As it explained, the shopping center, "by choice of its owner, is not limited to the personal use of appellants," but instead "is open to the public to come and go as they please."[341] In that setting, "[t]he views expressed by members of the public in passing out pamphlets or seeking signatures for a petition will not likely be identified with those of the owner."[342] The owner, moreover, remained free to distance itself from the message altogether. "As far as appears here," the Court noted, the shopping center could "expressly disavow any connection with the message by simply posting signs" disclaiming sponsorship.[343] Therefore, where third-party expression is neither attributable to the owner nor interferes with the owner's own speech, a requirement of access does not compel speech within the meaning of the First Amendment.[344]

## B. Implication

Once it is established that the Court protects editorial autonomy, understood as securing a speaker's control over what to publish, which subjects to address, how expressive materials are structured, the maintenance of expressive identity, and freedom from government-imposed speech, two implications naturally follow: First, when platforms select, filter, organize, rank, or deplatform content, whether AI-generated misinformation, sexually explicit material, or other sensitive or unpopular user expression, their decisions to host, tolerate, promote, demote, or exclude particular categories of content will be construed by courts as exercises of protected editorial discretion under the Free Speech Clause. State-enacted content moderation regulations, whether mandating or prohibiting moderation, are thus likely to be held unconstitutional when challenged. Second, corporations can no longer plausibly advance the claim that moderation requirements chill the free flow of information, as such arguments are unlikely to prevail.

---

[336] *Id* at 587.
[337] 477 U.S. 74 (1980).
[338] *Id* at 77.
[339] *Id.*
[340] *Id.*
[341] *Id* at 87.
[342] *Id.*
[343] *Id.*
[344] *Id.*

For the first implication, the claim that content moderation constitutes editorial judgment aligns squarely with the dominant school of internet governance scholarship. On this account, moderation is best understood as a privatized, hierarchical Weberian bureaucracy[345] that governs the inclusion, exclusion, and presentation of user-generated content.[346] Platforms review vast quantities of material posted by users and decide whether to retain or remove it.[347] To manage this scale, they adopt an industrial mode of governance,[348] deploying frontline reviewers, both human and automated,[349] to apply preexisting rules crafted by centralized policy teams to individual pieces of content. When errors occur, users may appeal, triggering additional rounds of review conducted within the same rule-bound framework.[350]

---

[345] MAX WEBER, ECONOMY AND SOCIETY: AN OUTLINE OF INTERPRETIVE SOCIOLOGY 957 (Guenther Roth & Claus Wittich eds., 1978) (describing bureaucracy as involving "a clearly established system of super- and subordination in which there is a supervision of the lower offices by higher ones," as well as "the possibility of appealing, in a precisely regulated manner, the decision of a lower office to the corresponding superior authority").

[346] *See* Kate Klonick, *The New Governors: The People, Rules, and Processes Governing Online Speech*, 131 HARV. L. REV. 1598, 1639-41 (2018). TARLETON GILLESPIE, CUSTODIANS OF THE INTERNET: PLATFORMSL, CONTENT MODERATION, AND THE HIDDEN DECISIONS THAT SHAPE SOCIAL MEDIA 5, 116 (2018). ("[P]latforms currently impose moderation at scale by turning some or all users into an identification force, employing a small group of outsourced workers to do the bulk of the review, and retaining for platform management the power to set the terms."); Kyle Langvardt, *Can the First Amendment Scale?,* 1 J. FREE SPEECH L. 273, 298 (2021) ("Legal culture's reflexive answer to these kinds of problems . . . is to require 'some kind of a hearing.' The 'hearing' may include confrontation rights, protective burdens of proof and production, opportunities for appeal, and so on . . . Many proposals to regulate or reform platform content moderation endorse this basic strategy, usually in combination with new transparency requirements." (footnotes omitted) (quoting Bd. of Regents v. Roth, 408 U.S. 564, 590 n.7 (1972))); Matthias C. Kettemann & Wolfgang Schulz, *Setting Rules for 2.7 Billion: A (First) Look into Facebook's Norm-Making System: Results of a Pilot Study 21–22* (Hans-Bredow-Institut, Paper No. 1, 2020), https://www.ssoar.info/ssoar/bitstream/handle/document/71724/ssoar-2020-kettemann_et_al-Setting_Rules_for_27_Billion.pdf?sequence=4&isAllowed=y&lnkname=ssoar-2020-kettemann_et_al-Setting_Rules_for_27_Billion.pdf (describing the "multi-step process" of rule development at Facebook, focusing on the Product Policy team which makes and changes the rules that are enforced by content moderators); ARTICLE 19, THE SOCIAL MEDIA COUNCILS: CONSULTATION PAPER 15 (2019), https://www.article19.org/wp-content/uploads/2019/06/A19-SMC-Consultation-paper-2019-v05.pdf (describing a proposal for a social media council that would sit above a platform's content moderation hierarchy and issue advisory opinions in individual cases); DAVID KAYE, SPEECH POLICE: THE GLOBAL STRUGGLE TO GOVERN THE INTERNET 53–57 (2019) (describing a "mini-legislative session" the author at- tended, which had the "vibe of a law school seminar" and involved questions of notice, due process, and appeal that are "exactly the right questions you would hope Facebook would be asking itself."); REBECCA MACKINNON, CONSENT OF THE NETWORKED: THE WORLDWIDE STRUGGLE FOR INTERNET FREEDOM 153–54 (2012) (describing the platform staff that "play the roles of lawmakers, judge, jury, and police all at the same time); Marvin Ammori, *The "New" New York Times: Free Speech Lawyering in the Age of Google and Twitter*, 127 HARV. L. REV. 2259, 2276 (2014) ("[T]he terms of service function much as traditional laws do": as rules "to be operationalized by hundreds of employees and contractors around the world . . . .").

[347] Kate Klonick, *The Facebook Oversight Board: Creating an Independent Institution to Adjudicate Online Free Expression*, 129 YALE L.J. 2418, 2427 (2020).

[348] *See* Gillespie, *supra* Note 346 at77.

[349] Elizabeth Dwoskin et al., *Content Moderators at YouTube, Facebook and Twitter See the Worst of the Web — and Suffer Silently*, WASHINGTON POST (July 25, 2019), https://www.washingtonpost.com/technology/2019/07/25/social-media-companies-are-outsourcing-their-dirty-work-philippines-generation-workers-is-paying-price/.

[350] *See* Kate Klonick, *The New Governors: The People, Rules, and Processes Governing Online Speech*, 131 HARV. L. REV. 1598, 1648 (2018)

The scale of this activity is immense. From July 2025 to September 2025 alone, YouTube removed more than seven million channels.[351] In the first quarter of 2022, TikTok removed over 102 million videos.[352] These figures do not include content that was reviewed and left in place, nor do they capture the volume of user appeals. It is therefore no exaggeration to say that the visible architecture of social media platforms is largely the product of the editorial judgments of these private governors. It follows that regulations targeting those judgments confront serious First Amendment obstacles.

Turning to the second implication, corporations can no longer plausibly maintain that mandatory moderation regimes chill the flow of information without asking the Court to embrace a chain of doctrinal moves it has increasingly refused to take. First, the Court would have to analogize social media platforms to "modern public squares,"[353] understood as the primary venues through which individuals communicate with family and friends, conduct business, perform daily tasks, and engage with public affairs, in the manner suggested by *Marsh*.[354] Second, it would have to treat restrictions on user-posted content as censorship, such that, under *Smith*[355] and *Bantam Books*,[356] both formal and informal regulatory pressures that risk chilling protected expression are constitutionally impermissible. Third, the Court would need to extend the fiduciary logic articulated in *Red Lion* to social media platforms.[357] None of these moves is likely to succeed, and taken together they are doctrinally implausible.

---

[351] See *YouTube Community Guidelines Enforcement*, GOOGLE TRANSPARENCY REPORT, https://transparencyreport.google.com/youtube-policy/removals?hl=en.

[352] TIKTOK, COMMUNITY GUIDELINES ENFORCEMENT REPORT (2022), https://www.tiktok.com/transparency/en-us/community-guidelines-enforcement-2022-3/

[353] *Moody*, 603 U.S at 767 (2024).

[354] 326 U.S. 501.

[355] 361 U.S. 147.

[356] 372 U.S. 58.

[357] Interestingly, this characterization of social media platforms as fiduciaries has also been advanced by Professor Jack Balkin since 2014, albeit on markedly different grounds. Professor Balkin argues that traditional First Amendment doctrine, developed for a communicative environment structured around speakers, listeners, and discrete acts of censorship, fails to apprehend the primary sites of power in a digital public sphere dominated by platforms and data-extractive infrastructures. As the practical conditions of speech increasingly depend on privately owned systems whose economic logic rests on data collection, inference, and behavioral prediction, end users become structurally dependent on these systems. Because this dependence is marked by significant informational asymmetries and heightened user vulnerability, Balkin contends that certain platforms should be treated as information fiduciaries, owing duties of care, confidentiality, and loyalty analogous to those long recognized in relationships such as doctor–patient or lawyer–client. *See* Jack M. Balkin, *Information Fiduciaries in the Digital Age*, BALKINIZATION (Mar. 5, 2014), https://balkin.blogspot.com/2014/03/information-fiduciaries-in-digital-age.html. He most fully elaborated his views in Jack M. Balkin, *Information Fiduciaries and the First Amendment*, 49 U.C. DAVIS L. REV. 1183 (2016). Additional discussions include Jack M. Balkin, Essay, *Free Speech in the Algorithmic Society: Big Data, Private Governance, and New School Speech Regulation*, 51 U.C. DAVIS L. REV. 1149, 1160–63 (2018); Jack M. Balkin, Essay, *Free Speech Is a Triangle*, 118 COLUM. L. REV. 2011, 2047–54 (2018); Jack M. Balkin & Jonathan Zittrain, *A Grand Bargain to Make Tech Companies Trustworthy*, The Atlantic (Oct. 3, 2016), https://www.theatlantic.com/technology/archive/2016/10/information-fiduciary/502346/. Professor Jonathan Zittrain has also been an important theorist and advocate of the information-fiduciary concept. *See, e.g.*, Balkin & Zittrain, supra; Jonathan Zittrain, *Facebook Could Decide an Election Without Anyone Ever Finding Out*, NEW REPUBLIC (June 1, 2014), https://newrepublic.com/article/117878/information-fiduciary-solution-facebook-digital-gerrymandering; Jonathan Zittrain, *How to Exercise the Power You Didn't Ask For*, HARV. BUS. REV. (Sept. 19, 2018), https://hbr.org/2018/09/how-to-exercise-the-power-you-didnt-ask-for; The academic literature taking up the idea of information fiduciaries has been overwhelmingly supportive. For representative responses from leading scholars of internet law, see Frank Pasquale, Lecture, *Response: Toward a Fourth Law of Robotics: Preserving Attribution, Responsibility, and*

With respect to the first two arguments, a closely related position was already advanced in *NetChoice, LLC v. Paxton*.[358] There, the Fifth Circuit upheld the Texas law by repeatedly invoking the premise that social media platforms function as "public squares,"[359] such that informal state pressure aimed at suppressing the circulation of disfavored material would violate the First Amendment. That framing, however, was never adopted by the Supreme Court. After a federal district court issued a preliminary injunction against the law and the Fifth Circuit stayed that injunction, the challengers sought emergency relief.[360] In a five-to-four order, the Court stayed the Fifth Circuit's decision and left the district court's injunction in place, preventing the Texas law from taking effect.[361] Chief Justice Roberts and Justices Breyer, Sotomayor, Kavanaugh, and Barrett formed the majority, though none issued an opinion,[362] while Justice Kagan dissented without explanation.[363] Even within the Fifth Circuit itself, subsequent precedent moved in the opposite direction. In *Missouri v. Biden*, the court emphasized that "social media platforms' content moderation decisions must be theirs and theirs alone," and held that it likely violates the First Amendment for the federal government to coerce or meaningfully encourage platforms to remove speech.[364]

The third argument fares no better. In *Red Lion Broadcasting Co.*,[365] the Court treated broadcast licensees as fiduciaries of the listening public, reasoning that spectrum scarcity, coupled with licensees' exclusive control over access to that public resource, created a structural power imbalance between speakers and audiences.[366] Because licensees could determine which views would be aired, excluded, or edited, the First Amendment permitted, and at times required, the imposition of special obligations designed to protect listeners' interests.[367]

Transposed to social media, the argument would run as follows: although the internet and data are not scarce, but instead defined by informational overabundance, platforms nonetheless exercise a comparable form of structural power over public discourse.[368] Through recommendation algorithms

---

*Explainability in an Algorithmic Society*, 78 OHIO ST. L.J. 1243, 1244 (2017) ("I believe that Balkin's concept of information fiduciary is well developed and hard to challenge."); and Tim Wu, Opinion, *An American Alternative to Europe's Privacy Law*, N.Y. TIMES (May 30, 2018), https://www.nytimes.com/2018/05/30/opinion/europe-america-privacy-gdpr.html ("[Technology] companies should be considered, to borrow a term coined by the law professor Jack Balkin, 'information fiduciaries' . . . ."). *But see contra.* Jane R. Bambauer, *The Relationships Between Speech and Conduct*, 49 U.C. DAVIS L. REV. 1941, 1949 (2016) (an "expansion of Balkin's proposal" to cover additional classes of data collectors, such as Netflix and Amazon, "could cause unsettling distortions of free speech protection.").
[358] 49 F.4th 439, 494 (5th Cir. 2022).
[359] *Id* at 46.
[360] *See* NetChoice, LLC v. Paxton, 142 S. Ct. 1715, 1715 (2022) (mem.). For an overview of the procedural complexity of the case, see id. at 1718 (Alito, J., dissenting).
[361] *Id* at 1715.
[362] *Id* at 1715-16 (without opinion or explanation)
[363] *See id.* at 1716 (Kagan, J., dissenting) (without opinion or explanation).
[364] No. 23-30445, 2023 WL 5821788 32 (5th Cir. Sept. 8, 2023) (citing Blum v. Yaretsky, 457 U.S. 991, 1008 (1982)).
[365] 395 U.S. 370-399.
[366] *Id.*
[367] *Id.*
[368] *See* Natali Helberger, *The Political Power of Platforms: How Current Attempts to Regulate Misinformation Amplify Opinion Power*, 8 DIGITAL JOURNALISM 842-854(2020) (arguing social media platforms to be political actors, wielding considerable opinion powers); *How Social Media Can Shape Public Opinion - and How We Can Build a Healthier Online Environment*,

and personalization systems, platforms selectively expose users to content that reinforces existing beliefs, producing echo chambers that functionally replicate the problem of one-sided access that animated *Red Lion*.[369] This imbalance is further aggravated by platforms' erosion of user autonomy,[370] intrusions upon privacy,[371] and systematic extraction and monetization of personal data.[372] On this view, social media platforms, like broadcast licensees, could be understood to stand in a fiduciary relationship to their audiences and to bear corresponding public obligations to ensure the presentation of competing viewpoints. If that premise were accepted, state laws restricting content moderation in order to promote viewpoint diversity would not offend the First Amendment.

Yet however intuitively appealing these arguments may sound, they fail as a matter of positive law. Platforms have prevailed in virtually every constitutional challenge to their content moderation decisions. Across dozens of cases contesting account suspensions, removals,[373] and deplatforming,[374] courts have almost uniformly held that moderation constitutes the enforcement of private editorial policies, not governmental censorship. Plaintiffs have repeatedly attempted to recharacterize platforms as state actors,[375] public forums,[376] common carriers,[377] or fiduciaries once they host user speech at scale or occupy a dominant position in the communications ecosystem. Those efforts have been consistently rejected. Neither market power nor daily use, nor even widespread use by public officials, transforms a private platform into a state actor for purposes of judicial review.

In this sense, the conclusion is straightforward. Because contemporary courts place decisive weight on editorial autonomy, and because platforms' curation practices are treated as exercises of editorial

---

GEORGETOWN UNIVERSITY (September 30, 2025), https://www.georgetown.edu/news/ask-a-professor-renee-diresta-how-social-media-can-shape-public-opinion/.
[369] *See* Peter Suciu, *Social Media Remains A Political Echo Chamber for the Likeminded*, FORBES (Jan 31, 2025), https://www.forbes.com/sites/petersuciu/2025/01/31/social-media-remains-a-political-echo-chamber-for-the-likeminded/.
[370] *See e.g.,* Ai Li, *The Illusion of "Community": How Red Note is Eroding User Autonomy with Personalization and Native Advertising*, MEDIUM (Sep 27, 2025), https://medium.com/@aili0178/the-illusion-of-community-how-red-note-is-eroding-user-autonomy-with-personalization-and-native-ca7447218747.
[371] *Supra* note 15.
[372] *Supra* note 118, ZUBOFF, *The Age of Surveillance Capitalism*.
[373] *See e.g.,* Domen v. Vimeo, Inc., 991 F.3d 66 (2d Cir. 2021); Federal Agency of News LLC v. Facebook, Inc., 432 F.Supp.3d 1107 (N.D. Cal. 2020); DeLima v. Google, Inc., 2021 WL 294560 (D.N.H.2021); Moates v. Facebook, Inc., 2024 WL 2853976 (N.D. Cal., 2024); Parler, LLC v. Amazon Web Services, Inc., 2021 WL 210721 (W.D. Wash. 2021); Enhanced Athlete, Inc. v. Google LLC, 479 F. Supp. 3d824 (N.D. Cal. 2020); Federal Agency of News LLC v. Facebook, Inc., 432 F.Supp.3d 1107 (N.D. Cal. 2020).
[374] *See e.g.,* Atkinson v. Facebook Inc., No. 20-cv-05546-RS (N.D. Cal. Dec.7, 2020); Elansari v. Jagex Inc., 790 Fed. Appx. 488 (3d Cir. 2020); Fyk v. Facebook, Inc., 808 F.Appx. 597 (9th Cir. 2020); Maffick LLC v. Facebook Inc., 2020 WL 5257853 (N.D. Cal. 2020); Davison v. Facebook, Inc., 370 F. Supp. 3d 621 (E.D. Va. 2019).
[375] *See* Manhattan Community Access Corp. v. Halleck, 587 U.S. _ (2019) (since the government either "compel[led] the private entity to take a particular action" or the "government act[ed] jointly with the private entity.")
[376] Moody, 603 U.S. 707
[377] *See* State v. Google LLC, No.21 CV H 06 0274 (Ohio Common Pleas Ct. complaint filed June 8, 2021), https://www.ohioattorneygeneral.gov/Files/Briefing-Room/News-Releases/Notice-of-Appeal-Time-Stamped.aspx (Ohio's Attorney General Dave Yost sued Google to establish "that Google's provision of internet search is properly classified as a common carrier and/or public utility under Ohio common law"); *see also* Biden v. Knight First Amendment Inst., 141 S.Ct. 1220 (2021) ("in many ways, digital platforms that hold themselves out to the public resemble traditional common carriers).

judgment, platforms' moderation decisions and the resulting modes of content presentation are likely to receive First Amendment protection. The harms inherent in such practices, to the extent they are legally cognizable at all, are left to private law, including tort doctrines such as strict liability where applicable, and to the liability framework established by Section 230. The First Amendment thus comes to be invoked primarily in service of outcomes that undermine the very constitutional order it was originally meant to sustain.

**PART III. Conclusion**

This Article begins with two types of AI-mediated harm: nonconsensual sexually explicit generative content and AI-generated political misinformation that destabilizes markets and undermines trust in institutions. Both present intuitively compelling cases for regulatory intervention aimed at preserving individual dignity and sustaining the conditions of democratic self-government. The Article then surveys state-level regulatory efforts pursuing these objectives, focusing primarily on measures directed at platforms. Against this background, the Article argues that, notwithstanding the seriousness and visibility of these harms, both AI-generated sexually explicit content and AI-generated political misinformation remain protected by the First Amendment under existing doctrine, for the following reasons: the obscenity doctrine does not extend to AI-generated images or fictional characters; hate speech is not categorically excluded from constitutional protection; and the fighting-words doctrine, as currently understood, does not plausibly include mediated, automated, or nonconfrontational AI-generated expression. As a result, content-based regulations targeting such material are unlikely to survive constitutional scrutiny.

The Article then turns to regulations that target content moderation. Via a close-reading of the Free Speech clause based on Founding-era mainstream dictionaries, and a mapping the Free Speech cases from the 1950s to the present, we find that the contemporary Free Speech doctrine has come to prioritize editorial autonomy, understood as control over authorship, selection, and expressive identity, together with a prohibition on state attribution of speech. Once, consistent with mainstream platform-governance scholarship, content moderation is characterized as an exercise of editorial judgment by platforms acting as private governors, it follows almost inexorably that regulations prohibiting or compelling such practices fall within the ambit of the Free Speech Clause.

The implication is straightforward: when the Free Speech Clause is invoked to invalidate regulations aimed at preserving individual autonomy and protecting children from manipulative and exploitative systems, it becomes detached from the purposes that once gave its protections public meaning. At that point, the Clause approaches a moment of doctrinal disintegration. If it is to continue serving its constitutional purpose of securing public participation in democratic self-government, it will require reconstruction into a coherent theory capable of addressing contemporary communicative infrastructures.